\documentclass[11pt,a4paper]{article}
\usepackage{amsmath,amssymb,amsfonts}
\usepackage[dvips]{graphicx}
\newtheorem{definition}{Definition}
\newtheorem{theorem}{Theorem}
\newtheorem{proposition}{Proposition}
\newtheorem{lemma}{Lemma}

\begin{document}
\title{Approximate resonance states in the semigroup decomposition of
resonance evolution}
\date{}
\maketitle
\begin{description}
 \item{$\ \ \ \ \ \ $}{\bf Y. Strauss\footnote{Electronic mail: ystrauss@cs.bgu.ac.il}}\\
        {\footnotesize\emph{Department of Mathematics, Ben-Gurion
        University of the Negev, Be'er Sheva 84105, Israel}}
 \item{$\ \ \ \ \ \ $}{\bf L.P. Horwitz\footnote{Electronic mail: larry@post.tau.ac.il}}\\
        {\footnotesize\emph{School of Physics, Raymond and Beverly
        Sackler Faculty of Exact Sciences,\\
        Tel-Aviv University, Ramat Aviv 69978, Israel\\
        and Physics Department, Bar-Ilan University, Ramat Gan, Israel\\
        and College of Judea and Samaria, Ariel, Israel}}
 \item{$\ \ \ \ \ \ $}{\bf A. Volovick\footnote{Electronic mail: volovyka@post.tau.ac.il}}\\
        {\footnotesize\emph{School of Physics, Raymond and Beverly
        Sackler Faculty of Exact Sciences,\\
        Tel-Aviv University, Ramat Aviv 69978, Israel}}
\end{description}
\bigskip
\begin{abstract}
  The semigroup decomposition formalism makes use of
  the functional model for $C_{\cdot 0}$
  class contractive semigroups for the description of the time
  evolution of resonances. For a given scattering problem the
  formalism allows for the association of a definite Hilbert space
  state with a scattering resonance. This state defines a
  decomposition of matrix elements of the evolution into a term
  evolving according to a semigroup law and a background term. We
  discuss the case of multiple resonances and give a bound on the size
  of the background term. As an example we treat a simple problem of scattering
  from a square barrier potential on the half-line.
\end{abstract}
\section{Introduction}
\label{intro}
\par Originally formulated for the analysis of scattering problems involving
solution of hyperbolic wave equations in the exterior domain of
compactly supported obstacles, the Lax-Phillips scattering theory${}^1$
was developed as a tool most suitable for dealing with resonances in
the scattering of electromagnetic or acoustic waves. Subsequent to its
introduction by Lax and Phillips, various
authors have contributed to further development of the theory${}^{2,3,4,5,6}$.
Notable recent additions were made by Sj\"ostrand and
Sworski${}^7$ who extended the scope of the theory to include general
classes of semibounded, compactly supported perturbations of the
Laplacian in the wave equation, and by Kuzhell, via the development of
a formalism providing conditions for the application of the
Lax-Phillips structure to an abstract form of the wave equation${}^8$
and to certain classes of Schr\"odinger operators${}^9$. 
In addition, Kuzhell and Moskalyova${}^{10}$ applied the Lax-Phillips
theory in the analysis of scattering systems involving singular
perturbations of the Laplacian. 
\par Several recent papers have dealt with the adaptation of the
Lax-Phillips theory to quantum mechanical scattering problems. An early work in this direction is
Ref. 11,12,13. A general formalism was developed in
Ref. 14 and subsequently applied 
to several physical models in Ref. 15,16,17. Such efforts to adapt the
Lax-Phillips formalism to the framework of quantum mechanics
are motivated by certain appealing features of the
Lax-Phillips theory. One of these features is the fact that the time evolution of
resonances in this theory is given in terms of a continuous, one parameter, strongly
contractive semigroup $\{\mathbf Z(t)\}_{t\geq 0}$
\begin{equation*}
 \mathbf Z(t_1)\mathbf Z(t_2)=\mathbf Z(t_1+t_2),\quad t_1,t_2\geq 0.
\end{equation*}
\par If $\mathcal H$ is a (separable) Hilbert space corresponding to a
particular scattering system and $\{\mathbf
U(t)\}_{t\in\mathbb R}$ is a unitary group defined on
$\mathcal H$ describing the evolution of the system, the basic
premises of the Lax-Phillips theory include the assumption of the
existence of an \emph{incoming subspace} $\mathcal D_-$ and an
\emph{outgoing subspace} $\mathcal D_+$ with respect to $\{\mathbf
U(t)\}_{t\in\mathbb R}$ which are assumed furthermore to be orthogonal
to each other. Denoting by $\mathbf P_-$ and $\mathbf P_+$ respectively the
projections on the orthogonal complements of $\mathcal D_-$ and
$\mathcal D_+$ in $\mathcal H$, and letting $\mathcal K=\mathcal H\ominus(\mathcal
D_-\oplus\mathcal D_+)$, the Lax-Phillips semigroup $\{\mathbf
Z(t)\}_{t\geq 0}$ defined by
\begin{equation}
\label{lp_semigroup}
 \mathbf Z(t)=\mathbf P_+\mathbf U(t)\mathbf P_-=\mathbf P_{\mathcal
 K}\mathbf U(t)\mathbf P_{\mathcal K},\quad t\geq 0,
\end{equation}
annihilates $\mathcal D_\pm$ and maps $\mathcal K$ into itself. 
The subspace $\mathcal K$ contains the scattering resonances and
the Lax-Phillips semigroup $\{\mathbf Z(t)\}_{t\geq 0}$ describes
their time evolution. In the Lax-Phillips framework resonances are
associated with pure states in the Hilbert space $\mathcal H$.
\par A basic difficulty encountered in the work on application of the
Lax-Phillips theory in quantum mechanics originates from the fact that
in this theory the continuous spectrum of the generator of evolution is required to
be unbounded from below as well as from above. Hence a formalism
utilizing the original structure of the theory, such as in Ref. 14,
is not suitable for application to large classes of
scattering problems in quantum mechanics (except for limited types of problems,
such as the Stark effect Hamiltonian${}^{17}$, or problems in a
relativistically covariant framework${}^{15,16}$, 
which can be analyzed by direct mapping to the Lax-Phillips structure.
The case of a Schr\"odinger equation with compactly supported potential may also
be analyzed within the Lax-Phillips framework through the use of the invariance
principle of wave operators${}^{18}$).
The subject of the present paper is a theoretical framework,
termed the \emph{semigroup decomposition} of resonance
evolution, developed with the goal of overcoming such difficulties.
Proposed by one of the authors (Y.S.) of the present
article${}^{19,20}$, this formalism makes use of the Sz.-Nagy-Foias
theory of contraction operators and contractive semigroups on Hilbert
space${}^{21}$ which, from the mathematical point of view, is the
fundamental theory underlying the Lax-Phillips construction through
the notion of model operators for $C_{\cdot 0}$ class semigroups 
(see Section \ref{semigroup_decomp_subsec}).
\par The presentation of the semigroup decomposition formalism in
Ref. 20 is based on the following assumptions:
\begin{enumerate}
\item[(i)] We are considering a scattering system consisting of a
      ``free'' unperturbed Hamiltonian $\mathbf H_0$ and a perturbed
      Hamiltonian $\mathbf H$, both defined on a Hilbert space
      $\mathcal H$.
\item[(ii)] $\mbox{ess\ supp}\,\sigma_{ac}(\mathbf H_0)=\mbox{ess\
        supp}\,\sigma_{ac}(\mathbf H)=\mathbb R^+$. For simplicity it
      is assumed further that the multiplicity of the a.c. spectrum is one.
\item[(iii)] The M\o ller wave operators $\mathbf\Omega^\pm\equiv\mathbf\Omega^\pm(\mathbf
      H_0,\mathbf H)$ exist and are complete.
\item[(iv)] The $S$-matrix in the \emph{energy representation} (the
      spectral representation for $\mathbf H_0$), denoted by
      $\tilde S(\cdot)$ has an extension to a meromorphic function
      $\mathcal S(\cdot)$ in an open, simply connected, region 
      $\Sigma\subset\mathbb C$ such that $\Sigma\cap\mathbb R$ is 
      an open interval in $\mathbb R$. The operator valued function $\mathcal S(\cdot)$ is
      holomorphic in $\Sigma\cap\mathbb C^+$ and has a simple
      pole (we generalize to the case of multiple poles in Section
      \ref{approx_res_state_sec} below) at a point 
      $z=\mu\in\Sigma\cap\mathbb C^-$ and no other
      singularity in $\overline\Sigma$, the closure of $\Sigma$.
\end{enumerate}
It is shown in Ref. 20 that there exists a dense set
$\Lambda\subset\mathcal H_{ac}(\mathbf H)$ and a well defined state 
$\psi_\mu\in\mathcal H_{ac}(\mathbf H)$ such that for any $g\in\Lambda$ and
any $f\in\mathcal H_{ac}$ the properties (i)-(iv) above induce, for
positive times, a decomposition of matrix elements of the
evolution $\mathbf U(t)$ in the form
\begin{equation}
\label{sem_dec_general_form}
 (g,\mathbf U(t)f)_{\mathcal H_{ac}(\mathbf
   H)}=R(g,f;t)+\alpha(g,\mu)(\psi_\mu,f)_{\mathcal H_{ac}(\mathbf H)}e^{-i\mu t},\quad t\geq 0.
\end{equation}
In a sense to be made precise in the next section the second term on
the right hand side of Eq. (\ref{sem_dec_general_form}) originates from an evolution semigroup
of Lax-Phillips type and the eigenvalue of the generator of this
semigroup is exactly $\mu$, i.e., the point of singularity of the
$S$-matrix. The quantity $R(g,f;t)$ on the right hand side of Eq. (\ref{sem_dec_general_form}) is
what we shall call a \emph{background} term. We note that if in Eq. (\ref{sem_dec_general_form}) we choose $f$
to be orthogonal to $\psi_\mu$ then the exponentially decaying semigroup term (second term on
the r.h.s. of Eq. (\ref{sem_dec_general_form}) ) vanishes. We call $\psi_\mu$ an \emph{approximate
 resonance state} and note that the characterization of $\psi_\mu$ as an
approximate resonance state rather than as an exact resonance state stems
from the fact that one can show (see Ref. 20) that there is no
choice of $g$ and $f$ that makes the backgound term $R(g,f;t)$ vanish.
\par An explicit expression for the approximate resonance state
$\psi_\mu$ is provided in Ref. 20. It is shown there that, if we denote by
$\{\vert E^-\rangle\}_{E\in\mathbb R^+}$ the set of outgoing solutions
of the Lippmann-Schwinger equation (using Dirac's notation), then $\psi_\mu$
is given by
\begin{equation}
\label{single_pole_approx_state}
 \psi_\mu=\frac{1}{2\pi i}\int_{\mathbb R^+}dE\,\frac{1}{E-\mu}\vert
 E^-\rangle .
\end{equation}
Following the introduction of approximate resonance states, 
the present paper discusses some generalizations. Thus, in
Section \ref{approx_res_state_sec} we assume that the region $\Sigma\cap\mathbb C^-$ contains
multiple resonance poles
of the $S$-matrix $\mathcal S(\cdot)$, say at $z=\mu_1,\ldots,\mu_n$
and obtain the form of the expression for the approximate resonance states
and semigroup decomposition of evolution matrix elements in this
case. In particular, we apply the semigroup decompostion to the
\emph{survival amplitude}, a central notion in the characterization of
the time evolution of resonances. Theorem \ref{background_estimate_thm} below then provides an
\emph{a priori} upper bound on the size of the background term in this case.
\par As a final remark we note that a modification of the Lax-Phillips
theory was recently used by H. Baumgartel for the description of
scattering resonances in certain quantum mechanical problems${}^{22}$ 
(see also Ref. 23). 
In particular, the assumption of orthogonality of $\mathcal D_\pm$,
essential in the context of the original Lax-Phillips formalism, is
replaced in Ref. 22 by
the requirement that an incoming subspace $\mathcal D_-$ and an outgoing
subspace $\mathcal D_+$ exist and the respective projections commute. The modified
assumptions on $\mathcal D_\pm$, accompanied by certain assumptions on 
$S$-matrix analyticity properties, result in a modified Lax-Phillips structure which
is then applied to the Friedrichs model, leading to the
construction of appropriate Gamow type vectors${}^{24}$ associated with scattering resonances.
The framework presented in Ref. 22 has several points of intersection
with the semigroup decomposition formalism discussed in the present
paper. The nature of these relationships will be discussed
elsewhere.
\par The rest of the paper is organized as follows: In 
Section \ref{semigroup_decomp_subsec} we describe the formalism
providing the semigroup decomposition of resonance evolution starting
with a short discussion of the functional model for $C_{\cdot 0}$
continuous contractive semigroups followed by a description of the
semigroup decomposition formalism introduced in Ref. 19,20. In
Section \ref{approx_res_state_sec} we extend the framework of 
Ref. 19,20 to the case of multiple resonances and,
furthermore, find an estimate on the size of the background term in
the expression for the time evolution of the survival probability of a
resonance. In Section \ref{example} we analyze a simple but illuminating example
involving a one dimensional model of scattering from a square barrier
potential. Section \ref{summary} contains a short summary of the
contents of the paper and some indication on further possible courses
of investigation.
\section{The semigroup decomposition for\\ resonance evolution}
\label{semigroup_decomp_subsec}
\subsection{Classification of contractive semigroups}
\par Several distinct classes of contractive semigroups are identified within the framework of the
Sz.-Nagy-Foias theory. Let $\{\mathbf T(t)\}_{t\geq 0}$ be a strongly
contractive semigroup defined on a Hilbert space $\mathcal H$. The classes $C_{0\cdot}$, $C_{\cdot 0}$,
$C_{1\cdot}$, $C_{\cdot 1}$ are defined by
\begin{eqnarray}
 \{\mathbf T(t)\}_{t\in\mathbb R^+}\in C_{0\cdot}\quad &\mbox{if}&
 \quad\mathbf T(t)h\to 0,\ \forall h\in\mathcal H\nonumber\\
 \{\mathbf T(t)\}_{t\in\mathbb R^+}\in C_{\cdot 0}\quad &\mbox{if}&
 \quad \mathbf T^*(t)h\to 0,\ \forall h\in\mathcal H\nonumber\\
 \{\mathbf T(t)\}_{t\in\mathbb R^+}\in C_{1\cdot}\quad &\mbox{if}&
 \quad \mathbf T(t)h\not\to 0,\ \forall h\in\mathcal H,\ h\not=0\nonumber\\
 \{\mathbf T(t)\}_{t\in\mathbb R^+}\in C_{\cdot 1}\quad &\mbox{if}&
 \quad \mathbf T^*(t)h\not\to 0,\ \forall h\in\mathcal H,\ h\not= 0\nonumber
\end{eqnarray}
The classes $C_{\alpha\beta}$ with $\alpha,\beta=0,1$ are then defined by
\begin{equation*}
 C_{\alpha\beta}=C_{\alpha\cdot}\cap C_{\cdot\beta},\qquad \alpha,\beta=0,1.
\end{equation*}
The semigroup $\{\mathbf Z(t)\}_{t\in\mathbb R^+}$ describing the time
evolution of resonances in the Lax-Phillips theory is readily
characterized by the fact that $\{\mathbf Z^*(t)\}_{t\in\mathbb R^+}$
belongs to the class $C_{\cdot 0}$\ . The structure of   
the Lax-Phillips outgoing spectral (and translation) representation is
then determined by that of the \emph{functional model}${}^{21,25}$ for $C_{\cdot
  0}$ class semigroups provided by the Sz.-Nagy-Foias theory. We say
an operator $\mathbf A$ is a \emph{model operator}${}^{25}$ for a given class
$C$ of operators if every operator in $C$ is similar to a multiple of
a part of $\mathbf A$ (a part of an operator $\mathbf A$ is a restriction of
$\mathbf A$ to one of its invariant subspaces). By a functional
model we mean that the model operator for a given class $C$ has a
canonical representation on suitable function spaces. For a
$C_{\cdot 0}$ class semigroup $\{\mathbf T(t)\}_{t\geq 0}$ the
associated functional model is essentially obtained through a
procedure of \emph{isometric dilation} of the \emph{cogenerator} of 
$\{\mathbf T(t)\}_{t\geq 0}$ and the similarity mapping to
the functional model is in fact a unitary transformation. 
\subsection{The functional model for $C_{\cdot 0}$ semigroups}
\label{functional_model_subsec}
\par We turn now to a brief description of the functional model for
semigroups in the class $C_{\cdot 0}$\ . Denote by $\mathbb C^+$ the upper half 
of the complex plane and let $H^2_{\mathcal N}(\mathbb C^+)$ be the 
Hardy space of vector valued functions analytic in
the upper half-plane and taking values in a separable Hilbert
space $\mathcal N$. The set of boundary values on $\mathbb
R$ of functions in $H^2_{\mathcal N}(\mathbb C^+)$, denoted below by
$H^2_{\mathcal N+}(\mathbb R)$, is a Hilbert space isomorphic to
$H^2_{\mathcal N}(\mathbb C^+)$. In a similar manner the Hardy space of
$\mathcal N$ valued functions analytic in the lower half-plane is
denoted by $H^2_\mathcal N(\mathbb C^-)$ and $H^2_{\mathcal
  N-}(\mathbb R)$ is the isomorphic Hilbert space consisting of
boundary values on $\mathbb R$ of functions in $H^2_{\mathcal
  N}(\mathbb C^-)$. Define $\{u(t)\}_{t\in\mathbb R}$, a family of unitary,
multiplicative operators $u(t): L^2_{\mathcal
  N}(\mathbb R)\mapsto L^2_{\mathcal N}(\mathbb R)$ by 
\begin{equation}
\label{translation_group_eqn}
 [u(t)f](\sigma)=e^{-i\sigma t}f(\sigma),\quad f\in L^2_{\mathcal
   N}(\mathbb R),\ \sigma\in\mathbb R.
\end{equation}
Assume that 
$\{\mathbf T(t)\}_{t\geq 0}$ is a $C_{\cdot 0}$ class
semigroup defined on a Hilbert space $\mathcal K$. Let the semigroup $\{\hat
T(t)\}_{t\geq 0}$, defined on a Hilbert space $\hat{\mathcal K}$, 
be the functional model for $\{\mathbf T(t)\}_{t\geq 0}$ and let $W:\mathcal
K\mapsto\hat{\mathcal K}$ be the similarity
transforming $\{\mathbf T(t)\}_{t\geq 0}$ into its functional model
$\{\hat T(t)\}_{t\geq 0}$ i.e., $\hat T(t)=W\mathbf T(t)W^{-1}$. Then
there exists a Hilbert space $\mathcal N$ such that 
$\hat{\mathcal K}$ is a closed subspace of $H^2_{\mathcal N +}(\mathbb R)$, $W$ is
unitary, and the functional model is given by
\begin{equation}
\label{cdotzero_functional_model_eqn}
 \hat T(t)= W\mathbf T(t)W^*=P_{\hat{\mathcal K}}u^*(t)\vert\hat{\mathcal K}
 ,\quad t\geq 0.
\end{equation}
Here $P_{\hat{\mathcal K}}$ is the orthogonal projection from
$H^2_{\mathcal N+}(\mathbb R)$ onto $\hat{\mathcal K}$, the subspace $\hat{\mathcal
K}$ is given by 
\begin{equation}
\label{cdotzero_compression_subspace_eqn}
 \hat{\mathcal K}=H^2_{\mathcal N +}(\mathbb
 R)\ominus\Theta_T(\cdot)H^2_{\mathcal N +}(\mathbb R),
\end{equation}
and $\Theta_T(\cdot): H^2_{\mathcal N +}(\mathbb R)\mapsto H^2_{\mathcal
  N +}(\mathbb R)$ is an \emph{inner function}${}^{26,29,30}$ for $H^2_{\mathcal N
  +}(\mathbb R)$ (depending, of course, on $\{\mathbf T(t)\}_{t\geq
  0}$) i.e., an operator valued function with the properties:
\begin{enumerate}
\item For each $\sigma\in\mathbb R$ the operator
      $\Theta_T(\sigma):\mathcal N\mapsto \mathcal N$ is the boundary value at $\sigma$
      of an operator valued function $\Theta_T(\cdot)$ analytic in the
      upper half-plane.
\item $\Vert\Theta_T(z)\Vert_{\mathcal N}\leq 1$ for ${\rm Im}\,z>0$.
\item $\Theta_T(\sigma)$, $\sigma\in\mathbb R$ is, pointwise, a
      unitary operator on $\mathcal N$.
\end{enumerate}
The operator valued function $\Theta_T(\cdot)$ is, in fact, the
\emph{characteristic function}${}^{21}$ of the cogenerator of the semigroup 
$\{\hat T(t)\}_{t\geq 0}$ (or $\{\mathbf T(t)\}_{t\geq 0}$). 
\par Let $P_+$ be the orthogonal projection of $L^2_{\mathcal N}(\mathbb
R)$ on $H^2_{\mathcal N +}(\mathbb R)$. The \emph{Toeplitz operator}
with symbol $u(t)$ (see, for example, Ref. 26,27 and references
therein), is an operator $T_{u(t)}: H^2_{\mathcal
  N +}(\mathbb R)\mapsto H^2_{\mathcal N +}(\mathbb R)$ defined by
\begin{equation}
\label{toeplitz_def_eqn}
 T_{u(t)}f\stackrel{def}{=}P_+u(t)f,\qquad f\in H^2_{\mathcal N
   +}(\mathbb R).
\end{equation}
We note that $\{T_{u(t)}\}_{t\geq 0}$ is a strongly contractive
semigroup on $H^2_{\mathcal N +}(\mathbb R)$ (see, for example,
Ref. 1,19,21,28. Taking the conjugate of $\hat T(t)$ in $H^2_{\mathcal
  N+}(\mathbb R)$ and using Eq. (\ref{cdotzero_functional_model_eqn}) 
one finds that
\begin{equation}
\label{cdotzero_star_functional_model_eqn}
 \hat T^*(t)= W\mathbf T^*(t)W^*=T_{u(t)}\vert\hat{\mathcal K},\quad
 t\geq 0.
\end{equation}
\par It follows from the discussion above that the Lax-Phillips semigroup 
$\{\mathbf Z(t)\}_{t\geq 0}$ has a functional model in the form of Eq.
(\ref{cdotzero_star_functional_model_eqn}) (recall that $\{\mathbf
Z^*(t)\}_{t\geq 0}$ is a $C_{\cdot 0}$ class semigroup), 
i.e., if we denote the functional model for $\{\mathbf
Z(t)\}_{t \geq 0}$ by $\{\hat Z(t)\}_{t\geq 0}$ then we have
\begin{equation}
\label{functional_model_eqn}
 \hat Z(t)=W\mathbf Z(t)W^*=T_{u(t)}\vert\hat{\mathcal K},\quad t\geq 0
\end{equation}
where $\hat{\mathcal K}\subset H^2_{\mathcal N+}(\mathbb R)$ is an
invariant subspace for $\{T_{u(t)}\}_{t\geq 0}$ given by
\begin{equation}
\label{compression_subspace_eqn}
 \hat{\mathcal K}=H^2_{\mathcal N+}(\mathbb R)\ominus
 \Theta_Z(\cdot)H^2_{\mathcal N+}(\mathbb R)
\end{equation}
and the inner function $\Theta_Z(\cdot)$ and the Hilbert space
$\mathcal N$ are determined by $\{\mathbf Z(t)\}_{t\geq 0}$.
A semigroup $\{\hat Z(t)\}_{t\geq 0}$ of the form given 
by Eq. (\ref{functional_model_eqn}) and
Eq. (\ref{compression_subspace_eqn}) 
is referred to in Ref. 20 as a \emph{Lax-Phillips type semigroup}. 
\par A central
theorem of the Lax-Phillips theory, corresponding to an important result
in the Sz.-Nagy-Foias theory relating the spectrum of a \emph{completely
non-unitary (cnu)} contraction to points of singularity of the
characteristic function states the following
\begin{theorem}
\label{semigroup_spectrum_thm}
Denote by $\hat B$ the generator of a Lax-Phillips type semigroup 
$\{\hat Z(t)\}_{t\geq 0}$. If ${\rm Im}\,\mu<0$, then $\mu$ belongs to
the point spectrum of $\hat B$ if and only if $\Theta^*_Z(\overline\mu)$ has a
nontrivial null space.
\end{theorem}
We note that the analytic continuation of $\Theta_Z(z)$ to the lower
half-plane is given by
\begin{equation*}
 \Theta_Z(z)\stackrel{def}{=}\left(\Theta^*_Z(\overline
   z)\right)^{-1},\qquad {\rm Im}\,z<0
\end{equation*}
and so a null space for $\Theta^*_Z(\overline\mu)$ implies the
existence of a pole for $\Theta_Z(z)$ at $z=\mu$. In the case of the
Lax-Phillips theory the characteristic function $\Theta_Z(\cdot)$ for
the Lax-Phillips semigroup is identical to the Lax-Phillips
$S$-matrix and its poles are the scattering resonances. As will be
seen below, the situation is a bit more involved in the semigroup
decomposition formalism.
\par We do not elaborate here further on the relations between
the functional model for $C_{\cdot 0}$ semigroups discussed above and 
the full structure of
the Lax-Phillips spectral representations and wave operators. The
reader is referred to Ref. 1,21. 
\subsection{The semigroup decomposition}
In order to apply the functional model for $C_{\cdot 0}$ semigroups ,
which is at the heart of the Lax-Phillips structure, to the
description of resonance evolution it is necessary to relate, for
$t\geq 0$, the evolution $\mathbf U(t)$ defined on the Hilbert space 
$\mathcal H$ of the scattering problem to the
Toeplitz evolution semigroup $T_{u(t)}$ of Eq. (\ref{toeplitz_def_eqn}) defined on 
$H^2_{\mathcal N +}(\mathbb R)$ and then restrict the latter,
according to Eq. (\ref{functional_model_eqn}), to a subspace
$\hat{\mathcal K}$ of $H^2_{\mathcal N
  +}(\mathbb R)$ associated with an appropriate inner function $\Theta_{\hat
  Z}(\cdot)$. In the framework of the Lax-Phillips theory this relation is 
guaranteed by the special properties of the Lax-Phillips incoming and
outgoing subspaces $\mathcal D_\pm$ (with $\mathcal D_-$ and $\mathcal
D_+$ denoting, respectively, the incoming and outgoing subspace), since in this case the
Lax-Phillips semigroup is a $C_{\cdot 0}$ semigroup. 
However, for many quantum mechanical scattering problems
one usually cannot find
subspaces with the properties of $\mathcal D_\pm$. A way of overcoming
this difficulty, proposed in Ref. 19 is to combine the standard functional
model for $C_{\cdot 0}$ semigroups with the notion of a
\emph{quasi-affine mapping} (see, for example, Ref. 21, Pg. 70):
\begin{definition}[Quasi-affine mapping]
A quasi-affine map from a Hilbert space $\mathcal H_1$ into a Hilbert
space $\mathcal H_0$ is a linear, one to one continuous mapping of $\mathcal H_1$
into a dense linear manifold in $\mathcal H_0$. If $\mathbf
A\in\mathcal B(\mathcal H_1)$ and $\mathbf B\in\mathcal B(\mathcal
H_0)$ then $\mathbf A$ is a quasi-affine transform of $\mathbf B$ if
there is a quasi-affine map $\theta: \mathcal H_1\mapsto\mathcal H_0$
such that $\theta\mathbf A=\mathbf B\theta$.
\end{definition}
\par\noindent The following theorem is proved in Ref. 19 for a scattering system 
consisting of unperturbed and perturbed Hamiltonians, respectively $\mathbf H_0$ and $\mathbf H$,
having semibounded continuous spectrum:
\begin{theorem}[Outgoing/Incoming contractive nesting]
\label{toeplitz_intertwining_thm}
Let $\mathbf H_0$ and $\mathbf H$ be self-adjoint operators on a Hilbert space
$\mathcal H$. Let $\{\mathbf U(t)\}_{t\in\mathbb R}$ be the unitary evolution group
on $\mathcal H$ generated by $\mathbf H$ {\rm [i.e, $\mathbf U(t)=
\exp(-i\mathbf H t)$]}. Denote by $\mathcal H_{ac}(\mathbf H_0)$ and
$\mathcal H_{ac}(\mathbf H)$, respectively, the absolutely continuous
subspaces of $\mathbf H_0$ and $\mathbf H$. Assume that the absolutely continuous 
spectrum of $\mathbf H_0$ and $\mathbf H$ has multiplicity one and that
$\mbox{ess\,Supp}\,\sigma_{ac}(\mathbf H_0)=\mbox{ess\,Supp}\,\sigma_{ac}(\mathbf H)
=\mathbb R^+$.
Assume furthermore that the M\o ller wave operators $\mathbf\Omega^\pm\equiv\mathbf\Omega^\pm(\mathbf H_0,
\mathbf H):\mathcal H_{ac}(\mathbf H_0)\mapsto\mathcal H_{ac}(\mathbf H)$
exist and are complete. Then there are mappings $\hat\Omega_\pm:
\mathcal H_{ac}(\mathbf H)\mapsto H^2_+(\mathbb R)$ such that
\begin{description}
\item[(i)] $\hat\Omega_\pm$ are contractive quasi-affine 
mappings of $\mathcal H_{ac}(\mathbf H)$ into $H^2_+(\mathbb R)$.
\item[(ii)] For every $t\geq 0$ the evolution $\mathbf U(t)$ is a
      quasi-affine transform of the Toeplitz operator $T_{u(t)}$ via
      the mapping $\hat\Omega_\pm$ i.e., for every
$f\in\mathcal H_{ac}(\mathbf H)$ we have
\begin{equation}
\label{transformed_evolution_eqn}
  \hat\Omega_\pm\mathbf U(t)f=T_{u(t)}\hat\Omega_\pm f\quad t\geq 0.
\end{equation}
\end{description}
\hfill$\square$
\end{theorem}
We call the triplet $(\mathcal H_{ac}(\mathbf H),H^2_+(\mathbb
R),\hat\Omega_-)$ the \emph{incoming contractive nesting} of
$\mathcal H_{ac}(\mathbf H)$ into $H^2_+(\mathbb R)$ and denote
$f_{in}=\hat\Omega_- f$. 
Similarly, the triplet $(\mathcal H_{ac}(\mathbf H),H^2_+(\mathbb R),\hat\Omega_+)$ 
is the \emph{outgoing contractive nesting} of $\mathcal H_{ac}(\mathbf
H)$ into $H^2_+(\mathbb R)$ and we denote $f_{out}=\hat\Omega_+ f$.
\par Define
\begin{equation*}
 \Xi_{\hat\Omega_+}\stackrel{{\mathrm{def}}}{=}\hat\Omega_+^*
 H^2_+(\mathbb R)\, .
\end{equation*}
Then, since $\hat\Omega_+^*$ is quasi-affine, the linear space
$\Xi_{\hat\Omega_+}\subset\mathcal H_{ac}(\mathbf H)$ is dense in 
$\mathcal H_{ac}(\mathbf H)$.
Moreover, since $\hat\Omega_+^*$ is one to one, for each
$g\in\Xi_{\hat\Omega_+}$ there is a unique $\tilde g\in H^2_+(\mathbb R)$
such that $g=\hat\Omega_+^*\tilde g$. We note that in Ref. 20
a dense set $\Lambda_{\hat\Omega_+}$, analogous to
$\Xi_{\hat\Omega_+}$, is defined somewhat differently, i.e.,
$\Lambda_{\hat\Omega_+}\stackrel{\rm def}{=}\hat\Omega_+^*\hat\Omega_+\mathcal H_{ac}(\mathbf
H)$. However, it will be seen below that the definition of $\Xi_{\hat\Omega_+}$ above, unlike
that of $\Lambda_{\hat\Omega_+}$, allows for a full characterization of
approximate resonance states. Using Theorem \ref{toeplitz_intertwining_thm}
we have, for every $g\in\Xi_{\hat\Omega_+}$ and $f\in\mathcal
H_{ac}(\mathbf H)$ and for $t\geq 0$
\begin{multline}
\label{hac_to_hardy_evolution}
 (g,\mathbf U(t)f)_{\mathcal H_{ac}(\mathbf H)}=(\hat\Omega_+^*\tilde
 g,\mathbf U(t)f)_{\mathcal H_{ac}(\mathbf H)}=\\
 =(\tilde g,T_{u(t)}\hat\Omega_+f)_{H^2_+(\mathbb R)}
 =(\tilde g,T_{u(t)}f_{out})_{H^2_+(\mathbb R)},\quad t\geq 0
\end{multline}
\par Following the definitions of the incoming and outgoing nestings of
$\mathcal H_{ac}(\mathbf H)$ into $H^2_+(\mathbb R)$ it is natural to define the
\emph{nested S-matrix}
$$ S_{nest}\stackrel{\mathrm{def}}{=}\hat\Omega_+\hat\Omega_-^{-1}\, .$$
Let $U: \mathcal H_{ac}(\mathbf H_0)\mapsto L^2(\mathbb R^+)$ be the
unitary transformation of $\mathcal H_{ac}(\mathbf H_0)$ onto the spectral
representation for $\mathbf H_0$ (also called the energy representation for $\mathbf H_0$).
If $\mathbf S=(\mathbf\Omega^-)^*\mathbf\Omega^+$ is the scattering operator
associated with $\mathbf H_0$ and $\mathbf H$ then
$\tilde S(\cdot):L^2(\mathbb R^+)\mapsto L^2(\mathbb R^+)$ defined by
\begin{equation}
\label{l2_s_matrix}
 \tilde S(\cdot)\stackrel{\mathrm{def}}{=}U\mathbf SU^*
\end{equation}
is the energy representation of the S-matrix. Let
$P_{\mathbb R^+}: L^2(\mathbb R)\mapsto L^2(\mathbb R)$ be the orthogonal
projection in $L^2(\mathbb R)$ on the subspace of functions supported on
$\mathbb R^+$ and define the inclusion map $I: L^2(\mathbb R^+)\mapsto
L^2(\mathbb R)$ by
\begin{equation}
\label{l2_l2_inclusion}
 (If)(\sigma)=\begin{cases}
                 f(\sigma), & \sigma\geq 0\\
                 0         ,& \sigma<0\, .\\
              \end{cases}
\end{equation}
Then the inverse $I^{-1}: P_{\mathbb R^+}L^2(\mathbb R)\mapsto L^2(\mathbb R^+)$
is, of course, one to one on $P_{\mathbb R^+}L^2(\mathbb R)$.
Let $\theta: H^2_+(\mathbb R)\mapsto L^2(\mathbb R^+)$ be a map
given by
\begin{equation}
\label{theta_map}
 \theta f=I^{-1}P_{\mathbb R_+}f,\quad f\in H^2_+(\mathbb R)\, .
\end{equation}
By a theorem of Van Winter${}^{31}$, $\theta$ is a quasi-affine
transform mapping $H^2_+(\mathbb R)$ into $L^2(\mathbb R^+)$. The adjoint
map $\theta^*: L^2(\mathbb R^+)\mapsto H^2_+(\mathbb R)$ is then also a
contractive quasi-affine map. An explicit expression for $\theta^*$ is
provided by the following lemma${}^{19}$:
\begin{lemma}
\label{theta_star_lemma}
Let $I: L^2(\mathbb R^+)\mapsto L^2(\mathbb R)$ be the inclusion map defined
in Eq. (\ref{l2_l2_inclusion}). Let $P_+$ be the orthogonal projection of
$L^2(\mathbb R)$ onto $H^2_+(\mathbb R)$. Then for every $f\in L^2(\mathbb R^+)$
we have
\begin{equation}
\label{theta_star_explicit}
 \theta^*f=P_+ If,\quad f\in L^2(\mathbb R^+)\, .
\end{equation}
\hfill$\square$
\end{lemma}
It is shown in Ref. 19 that the nested S-matrix can be
expressed in the form 
\begin{equation}
\label{s_nest_explicit}
 S_{nest}=\theta^*\tilde S(\cdot)(\theta^*)^{-1}\, .
\end{equation}
\par Following Ref. 20 we now use assumption (iv) in Section
\ref{intro}. The S-matrix $\tilde S(\cdot)$ is then the restriction of its extension
$\mathcal S(\cdot)$ on $\mathbb R^+$. Under these assumptions 
$\mathcal S(\cdot)$ has, in the region $\Sigma$, a representation of
the form (see Ref. 20)
\begin{equation*}
 \mathcal S(z)=\mathcal B_\mu(z)\mathcal S'(z),\quad z\in\Sigma
\end{equation*}
where
\begin{equation}
\label{blaschke_factor}
 \mathcal B_\mu(z)=\frac{z-\overline\mu}{z-\mu},\quad
 z\in\mathbb C\backslash \{\mu\}
\end{equation}
and $\mathcal S'(\cdot)$ is analytic and has no zeros in $\Sigma$.
Restricting $\mathcal S(\cdot)$ to the positive real axis we obtain
\begin{equation}
\label{s_tilde_factorization_eqn}
 \tilde S(E)=\tilde B_\mu(E)\tilde S'(E),\quad E\geq 0\, 
\end{equation}
where by definition
$\tilde B_\mu(E)\stackrel{def}{=}\mathcal B_\mu(E)$ and 
$\tilde S'(E)\stackrel{def}{=}\mathcal S'(E)$ for $E\geq 0$.
We note that both $\tilde S'(\cdot)$ and $\tilde B_\mu(\cdot)$ are considered
here as multiplicative unitary operators on $L^2(\mathbb R^+)$ (moreover, they
are pointwise unitary a.e. for $E\geq 0$). Moreover, $\mathcal B_\mu(\cdot)$ can
be regarded as a multiplicative operator on $L^2(\mathbb R)$. In fact,
considered as a multiplicative
operator on $H^2_+(\mathbb R)\subset L^2(\mathbb R)$, 
$\mathcal B_\mu$ is a \emph{Blaschke} factor (the
definition of Blaschke products and Blaschke factors can be found, for example
in Ref. 29,30, see e.g., Eq. (\ref{multi_resonance_blaschke_eqn})
below). Such a factor is the simplest example of an inner function
for $H^2_+(\mathbb R)$. We make use of this fact through the following
proposition, not stated as such, but implicitly used in Ref. 20:
\begin{proposition}
\label{blaschke_decomp_prop}
Let $\tilde B_\mu(\cdot): L^2(\mathbb R^+)\mapsto L^2(\mathbb R^+)$ be defined by
$\tilde B_\mu(E)=\mathcal B_\mu(E)$, $E\geq 0$ where $\mathcal B_\mu(\cdot)$ is
defined in Eq. (\ref{blaschke_factor}). Let $\theta^*: L^2(\mathbb R^+)
\mapsto H^2_+(\mathbb R)$ be the adjoint of the map $\theta$ defined in
Eq. (\ref{theta_map}). Let $\hat{\mathcal K}_\mu\subset H^2_+(\mathbb
R)$ and $\hat{\mathcal K}_{\overline\mu}\subset H^2_-(\mathbb R)$ be subspaces
defined by
\begin{equation*}
 \hat{\mathcal K}_\mu\stackrel{def}{=}H^2_+(\mathbb R)\ominus\mathcal
 B_\mu(\cdot)H^2_+(\mathbb R),
 \qquad 
 \hat{\mathcal K}_{\overline\mu}\stackrel{def}{=}H^2_-(\mathbb R)\ominus\mathcal
 B_{\overline\mu}(\cdot)H^2_-(\mathbb R)\, ,
\end{equation*}
where $\mathcal B_{\overline\mu}(z)=(z-\mu)(z-\overline\mu)^{-1}$ and
denote by $P_{\hat{\mathcal K}_\mu}$ and
$P_{\hat{\mathcal K}_{\overline\mu}}$ the orthogonal projections of
$L^2(\mathbb R)$ on
$\hat{\mathcal K}_\mu$ and $\hat{\mathcal K}_{\overline\mu}$ respectively.
For every $f\in L^2(\mathbb R^+)$ we then have
\begin{equation}
\label{blaschke_decomp_eqn}
 \theta^*\tilde B_\mu f=\mathcal B_\mu\theta^* f
 +P_{\hat{\mathcal K}_\mu}\mathcal B_\mu P_{\hat{\mathcal K}_{\overline\mu}}
 \overline\theta^*f\, .
\end{equation}
here $\overline\theta^*: L^2(\mathbb R^+)\mapsto H^2_-(\mathbb R)$ and 
$\overline\theta^* f=P_-If$ with $I$ defined in Eq. (\ref{l2_l2_inclusion}) and
$P_-$ the orthogonal projection of $L^2(\mathbb R)$ onto $H^2_-(\mathbb R)$.
\hfill$\square$
\end{proposition}
\par{\bf Proof:} Using Eq. (\ref{theta_star_explicit}) in Lemma
\ref{theta_star_lemma} we get
\begin{equation*}
 \theta^*\tilde B_\mu f=P_+I\tilde B_\mu f=P_+\mathcal B_\mu If=
 P_+\mathcal B_\mu (P_+ +P_-)If
 =P_+\mathcal B_\mu \theta^* f+P_+\mathcal B_\mu P_-{\overline\theta}^*f
\end{equation*}
Eq. (\ref{blaschke_decomp_eqn}) then follows from the fact, proved in
Ref. 20, that $P_+\mathcal B_\mu P_-=P_{\hat{\mathcal K}_\mu}\mathcal B_\mu
P_{\hat{\mathcal K}_{\overline\mu}}$ and from the property of $\mathcal B_\mu(\cdot)$
of being an inner function for $H^2_+(\mathbb R)$.
\par\hfill$\blacksquare$
\bigskip
\par We note that since $\mathcal B_\mu(\cdot)$ is an inner function 
Eq. (\ref{functional_model_eqn}), (\ref{compression_subspace_eqn}) and
Theorem \ref{semigroup_spectrum_thm} imply that
\begin{equation}
\label{compressed_evolution_eqn}
 T_{u(t)}P_{\hat{\mathcal K}_\mu}=\hat Z(t)P_{\hat{\mathcal K}_\mu}
 =e^{-i\mu t}P_{\hat{\mathcal K}_\mu},\quad t\geq 0.
\end{equation}
Combining Eq. (\ref{s_nest_explicit}),
Eq. (\ref{s_tilde_factorization_eqn}) and Eq. (\ref{blaschke_decomp_eqn}) we
obtain
\begin{multline}
\label{f_out_decomp_eqn}
 f_{out}=S_{nest}f_{in}=\theta^*\tilde S(\theta^*)^{-1}f_{in}
 =\theta^*\tilde B_\mu\tilde S'(\theta^*)^{-1}f_{in}=\\
 =\mathcal B_\mu\theta^*\tilde S'(\theta^*)^{-1}f_{in}
 +P_{\hat{\mathcal K}_\mu}\mathcal B_\mu P_{\hat{\mathcal K}_{\overline\mu}}
 {\overline\theta^*}\tilde S'(\theta^*)^{-1}f_{in}\, .$$
\end{multline}
Using the decomposition of $f_{out}$ from Eq. (\ref{f_out_decomp_eqn}) in the
r.h.s. of Eq. (\ref{hac_to_hardy_evolution}) and applying 
Eq. (\ref{compressed_evolution_eqn}) we
obtain the \emph{semigroup decomposition} for $t\geq 0$ of the time
evolution corrsponding to the resonance at $z=\mu$

\begin{multline}
\label{semigroup_decomp}
 (g,\mathbf U(t)f)_{\mathcal H_{ac}(\mathbf H)}
 =(\tilde g,T_{u(t)}f_{out})_{H^2_+(\mathbb R)}=\\
 =(\tilde g,u(t)\mathcal B_\mu\theta^*\tilde S'(\theta^*)^{-1}f_{in})_{H^2_+(\mathbb R)}
 +e^{-i\mu t}(\tilde g, \mathcal B_\mu {\overline\theta^*}\tilde S'
 (\theta^*)^{-1}f_{in})_{H^2_+(\mathbb R)}\, .
\end{multline}
As is seen above in Eq. (\ref{compressed_evolution_eqn}), the exponential decay in the second term on the r.h.s of
Eq. (\ref{semigroup_decomp}) originates with the semigroup $\hat Z(t)$. The first term
on the r.h.s. of Eq. (\ref{semigroup_decomp}) is the
\emph{background term} and is responsible for deviations from a purely
exponential decay law.
\section{Approximate resonance states}
\label{approx_res_state_sec}
\par It is an interesting fact that the semigroup decomposition described in
the previous section associates a unique state in $\mathcal H_{ac}(\mathbf H)$
with a resonance pole at $z=\mu$ (${\rm Im}\,\mu<0$). The following
theorem is proved in Ref. 20
\begin{theorem}[approximate resonance state]
\label{approx_resonance_state_thm}
Under the assumptions of Theorem \ref{toeplitz_intertwining_thm}, let
$\tilde S: L^2(\mathbb R^+)\mapsto L^2(\mathbb R^+)$ be the S-matrix in the
energy representation defined in Eq. (\ref{l2_s_matrix}). Assume that
$\tilde S(\cdot)$ is the restriction to $\mathbb R^+$ of a function $\mathcal S(\cdot)$
meromorphic in an open region $\Sigma$ with a single, simple pole at a point
$z=\mu$, $\mu\in\Sigma\cap\mathbb C^-$. For any $f\in\mathcal H_{ac}(\mathbf H)$
define $f_{out}=\hat\Omega_+f$ and $f_{in}=\hat\Omega_-f$. There exists a
unique state $\psi_\mu\in\mathcal H_{ac}(\mathbf H)$ such that
\begin{equation}
\label{approx_state_decomp_eqn}
 f_{out}=\mathcal B_\mu\theta^*\tilde S'(\theta^*)^{-1}f_{in}
 +\frac{\vert{\rm Im}\,\mu\vert}{\pi}(\psi_\mu,f)_{\mathcal H_{ac}(\mathbf H)}
 \,x_\mu
\end{equation}
where $\theta^*$ is the map given by lemma
\ref{theta_star_lemma}, $\mathcal B_\mu$ is given in Eq. (\ref{blaschke_factor}),
$\tilde S'$ is defined by Eq. (\ref{s_tilde_factorization_eqn}) and
$x_\mu\in H^2_+(\mathbb R)$ is given by
$x_\mu(\sigma)=(\sigma-\mu)^{-1}$, $\sigma\in\mathbb R$.\hfill$\square$
\end{theorem}
Combining Eq. (\ref{approx_state_decomp_eqn}) and Eq. (\ref{semigroup_decomp})
we can write the semigroup decomposition in the form
\begin{multline}
\label{semigroup_decomp_approx_state_eqn}
 (g,\mathbf U(t)f)_{\mathcal H_{ac}(\mathbf H)}
 =(\tilde g,u(t)\mathcal B_\mu\theta^*
 \tilde S'(\theta^*)^{-1}f_{in})_{H^2_+(\mathbb R)}\\
 +\frac{{\rm Im}\,\mu}{\pi}\,e^{-i\mu t}(\tilde g,x_\mu)_{H^2_+(\mathbb R)}
 (\psi_\mu,f)_{\mathcal H_{ac}(\mathbf H)},\quad t\geq 0
\end{multline}
where $g\in\Xi_{\hat\Omega_+}$ and $\tilde g=(\hat\Omega_+^*)^{-1}g$.
The eigenstate $x_\mu\in H^2_+(\mathbb R)$ of the semigroup $\hat Z(t)$ providing
the exponential decay of the second term on the r.h.s. of
Eq. (\ref{semigroup_decomp_approx_state_eqn}) is called below the \emph{Hardy space
resonance state}. The state $\psi_\mu\in \mathcal H_{ac}(\mathbf H)$ whose
existence is implied by Theorem \ref{approx_resonance_state_thm} is called
\emph{approximate resonance state}. We observe that if in
Eq. (\ref{semigroup_decomp_approx_state_eqn}) we choose $f\in\mathcal
H_{ac}(\mathbf H)$ orthogonal to $\psi_\mu$ then the second term on
the r.h.s. of that equation is identically zero.
\par Denote by $\{\vert E^-\rangle\}_{E\in\mathbb R^+}$ the set
of outgoing solutions of the Lippmann-Schwinger equation. For every
$f\in\mathcal H_{ac}(\mathbf H)$ we have
\begin{equation}
\label{outgoing_lipmann_schwinger_eqn}
 (U(\mathbf\Omega^-)^*f)(E)=\langle E^-\vert f\rangle,\quad E\in\mathbb R^+
\end{equation}
It is shown in Ref. 20 that an explicit expression for the
approximate resonance state $\psi_\mu$ is given by
\begin{equation}
\label{approx_resonance_state_explicit_eqn}
 \psi_\mu=\frac{1}{2\pi i}\int_{\mathbb R^+}dE\,\frac{1}{E-\mu}\vert E^-\rangle\, .
\end{equation}
\par In this section we explore several properties of approximate resonance
states $\psi_\mu$. Our first step is to extend the discussion above to the
case of multiple resonances:
\begin{theorem}[multiple resonance case]
\label{multiple_resonance_thm}
Under the assumptions of Theorem \ref{toeplitz_intertwining_thm},
let $\tilde S: L^2(\mathbb R^+)\mapsto L^2(\mathbb R^+)$ be the $S$-matrix in
the energy representation defined in Eq. (\ref{l2_s_matrix}). Assume that
$\tilde S(\cdot)$ is the restriction to $\mathbb R^+$ of a function $\mathcal S(\cdot)$
meromorphic in the open region $\Sigma$ with $n$ simple poles at points
$z=\mu_i$, $i=1,\ldots,n$, $\mu_i\in\Sigma\cap\mathbb C^-$. Then there exist
$n$ distinct states $\{\psi^\Sigma_{\mu_i}\}_{i=1,\ldots,n}$,
$\psi^\Sigma_{\mu_i}\in \mathcal H_{ac}(\mathbf H)$, such that for every
$f\in\mathcal H_{ac}(\mathbf H)$ we have
\begin{equation}
\label{multi_state_sem_dec_eqn}
 f_{out}=\mathcal B_{\mu_1\ldots\mu_n}\theta^*\tilde S'(\theta^*)^{-1}f_{in}
 +\sum_{j=1}^n\frac{\vert{\rm Im}\,\mu_j\vert}{\pi}
 \prod_{\substack{i=1\\ i\not=j}}^n
 \frac{\mu_j-\overline\mu_i}{\mu_j-\mu_i}
 (\psi^\Sigma_{\mu_j},f)_{\mathcal H_{ac}(\mathbf H)}\,x_{\mu_j}
\end{equation}
where
\begin{equation}
\label{multi_resonance_blaschke_eqn}
 \mathcal B_{\mu_1\ldots\mu_n}(z)\stackrel{\mathrm{def}}{=}
 \prod_{i=1}^n \frac{z-\overline\mu_i}
 {z-\mu_i}\,.
\end{equation}
In Eq. (\ref{multi_state_sem_dec_eqn}) $\tilde S'(\cdot)$ is the restriction
to $\mathbb R^+$ of a function $\mathcal S'(\cdot)$ analytic in $\Sigma$ and
having no poles in $\overline\Sigma$ and $x_{\mu_j}(\sigma)=(\sigma-\mu_j)^{-1}$,
$\sigma\in\mathbb R$. The states $\psi^\Sigma_{\mu_j}$, $j=1,\ldots,n$ are given by
\begin{equation}
\label{multiple_resonance_psi_mu_eqn}
 \psi^\Sigma_{\mu_j}=\int_{\mathbb R^+} dE
 \prod_{\substack{i=1\\i\not=j}}^n\frac{E-\overline\mu_i}{E-\mu_i}
 \frac{1}{E-\mu_j}\vert E^-\rangle\, . 
\end{equation}
\break ${\, }$\hfill$\square$
\end{theorem}
\par{\bf Proof:} Assume that $\mathcal S(\cdot)$, the extension of $\tilde S(\cdot)$ from
$\mathbb R^+$ into $\Sigma\cup\mathbb R^+$ has $n$ simple poles in
$\Sigma\cap\mathbb C^-$. Then, applying the same arguments as in
Ref. 20, we find that $\mathcal S(\cdot)$ can be factorized in
$\Sigma$ in the form
\begin{equation*}
 \mathcal S(z)=\mathcal B_{\mu_1\ldots\mu_n}(z)\mathcal S'(z)
\end{equation*}
where $\mathcal B_{\mu_1\ldots\mu_n}$, defined in
Eq. (\ref{multi_resonance_blaschke_eqn}), is a finite Blaschke product and
$\mathcal S'(\cdot)$ has no poles in $\Sigma$. In addition we have, of
course
\begin{equation*}
 \tilde S(E)=\tilde B_{\mu_1\ldots\mu_n}(E)\tilde S'(E),\quad E\geq
 0\, .
\end{equation*}
The semigroup decomposition then follows exactly as in Section \ref{intro}
with $\mathcal B_{\mu_1\ldots\mu_n}$ and $\tilde B_{\mu_1\ldots\mu_n}$
replacing $\mathcal B_\mu$ and $\tilde B_\mu$ respectively. For the resonance
term in Eq. (\ref{f_out_decomp_eqn}) we get in this case 
\begin{equation*}
 P_+\mathcal B_{\mu_1\ldots\mu_n}P_-\overline\theta^*\tilde S'(\theta^*)^{-1}
 f_{in},\quad f\in\mathcal H_{ac}(\mathbf H)
\end{equation*}
Recalling that
\begin{equation*}
 (P_+f)(\sigma)=\frac{1}{2\pi i}\int_{-\infty}^\infty d\sigma'\,
 \frac{1}{\sigma-\sigma'+i0}f(\sigma'),
 \quad f\in L^2(\mathbb R),\ \ \sigma\in\mathbb R
\end{equation*}
and
\begin{equation*}
 (\overline\theta^* f)(\sigma)=\frac{1}{2\pi i}
 \int_0^\infty dE\,\frac{1}{E-\sigma+i0} f(E),
 \qquad f\in L^2(\mathbb R^+)
\end{equation*}
(see Ref. 19) we obtain
\begin{multline*}
 (P_+\mathcal B_{\mu_1\ldots\mu_n}P_-\overline\theta^*
 \tilde S'(\theta^*)^{-1}f_{in})(\sigma)=\\
 =\frac{-1}{4\pi^2}\int_0^\infty dE\int_{-\infty}^\infty d\sigma'
 \frac{1}{\sigma-\sigma'+i0}\prod_{i=1}^n \frac{\sigma'-\overline\mu_i}
 {\sigma'-\mu_i}\frac{1}{E-\sigma'+i0}\tilde S'(E)((\theta^*)^{-1}f_{in})(E)=\\
 =\sum_{j=1}^n\frac{\vert{\rm Im}\,\mu_j\vert}{\pi}\frac{1}{\sigma-\mu_j}
 \int_0^\infty dE \prod_{\substack{i=1\\ i\not=j}}^n
 \frac{\mu_j-\overline\mu_i}
 {\mu_j-\mu_i}\frac{1}{E-\mu_j}\tilde S'(E)((\theta^*)^{-1}f_{in})(E)=\\
 =\sum_{j=1}^n\frac{\vert{\rm Im}\,\mu_j\vert}{\pi}\,x_{\mu_j}(\sigma)
 \prod_{\substack{i=1\\ i\not=j}}^n\frac{\mu_j-\overline\mu_i}{\mu_j-\mu_i}
 \int_0^\infty dE \frac{1}{E-\overline\mu_j}
 \prod_{\substack{i=1\\i\not=j}}^n\frac{E-\mu_i}{E-\overline\mu_i}
 \tilde S(E)((\theta^*)^{-1}f_{in})(E)=\\
 =\sum_{j=1}^n\frac{\vert{\rm Im}\,\mu_j\vert}{\pi}\,x_{\mu_j}(\sigma)
 \prod_{\substack{i=1\\ i\not=j}}^n\frac{\mu_j-\overline\mu_i}{\mu_j-\mu_i}
 \int_0^\infty dE \frac{1}{E-\overline\mu_j}
 \prod_{\substack{i=1\\i\not=j}}^n\frac{E-\mu_i}{E-\overline\mu_i}
 (U(\mathbf\Omega^-)^* f)(E)=\displaybreak[2]\\
 =\sum_{j=1}^n\frac{\vert{\rm Im}\,\mu_j\vert}{\pi}\,x_{\mu_j}(\sigma)
 \prod_{\substack{i=1\\ i\not=j}}^n\frac{\mu_j-\overline\mu_i}{\mu_j-\mu_i}
 \int_0^\infty dE \frac{1}{E-\overline\mu_j}
 \prod_{\substack{i=1\\i\not=j}}^n\frac{E-\mu_i}{E-\overline\mu_i}
 \langle E^-\vert f\rangle\\
\end{multline*}
where $x_{\mu_j}\in H^2_+(\mathbb R)$ is the Hardy space resonance state
corresponding to $\mu_j$ i.e., $x_{\mu_j}(\sigma)=(\sigma-\mu_j)^{-1}$.
Defining the states $\psi^\Sigma_{\mu_j}$, $j=1,\ldots n$ according to
Eq. (\ref{multiple_resonance_psi_mu_eqn}) we obtain
\begin{equation}
\label{psi_mu_expansion_eqn}
 (P_+\mathcal B_{\mu_1\ldots\mu_n}P_-\overline\theta^*
 \tilde S'(\theta^*)^{-1}f_{in})(\sigma)
 =\sum_{j=1}^n\frac{\vert{\rm Im}\,\mu_j\vert}{\pi}\,
 \prod_{\substack{i=1\\ i\not=j}}^n\frac{\mu_j-\overline\mu_i}{\mu_j-\mu_i}
 (\psi^\Sigma_{\mu_j},f)_{\mathcal H_{ac}(\mathbf H)}\,x_{\mu_j}(\sigma)
\end{equation}
This proves Theorem \ref{multiple_resonance_thm}.\hfill$\blacksquare$
\bigskip
\par\noindent We observe that Eq. (\ref{multiple_resonance_psi_mu_eqn}) is a generalization
of Eq. (\ref{approx_resonance_state_explicit_eqn}). Hence $\psi^\Sigma_{\mu_j}$ is
the approximate resonance state corresponding to the pole of
$\mathcal S(\cdot)$ at $z=\mu_j$. Combining Eq. (\ref{psi_mu_expansion_eqn})
and Eq. (\ref{semigroup_decomp}) we get the semigroup decomposition for the
multi-resonance case
\begin{multline}
\label{sem_dec_multi_resonance_eqn}
 (g,\mathbf U(t)f)_{\mathcal H_{ac}(\mathbf H)}
 =(\tilde g,u(t)\mathcal B_{\mu_1\ldots\mu_n}\theta^*
 \tilde S'(\theta^*)^{-1}f_{in})_{H^2_+(\mathbb R)}\\
 +\sum_{j=1}^n\frac{\vert{\rm Im}\,\mu_j\vert}{\pi}\,
 \prod_{\substack{i=1\\ i\not=j}}^n\frac{\mu_j-\overline\mu_i}{\mu_j-\mu_i}
 (\psi^\Sigma_{\mu_j},f)_{\mathcal H_{ac}(\mathbf H)}\,
 (\tilde g,x_{\mu_j})_{H^2_+(\mathbb R)}e^{-i\mu_j t},\quad t\geq 0\,.
\end{multline}
\par The approximate resonance states in Eq. (\ref{multiple_resonance_psi_mu_eqn})
and semigroup decomposition of Eq. (\ref{multi_state_sem_dec_eqn}) and Eq.
(\ref{sem_dec_multi_resonance_eqn}) depend, of course, on the region $\Sigma$.
If $\{\mu_j\}_{j=1,\ldots,n}$ are the poles in $\Sigma\cap\mathbb C^-$
of the meromorphic extension
$\mathcal S(\cdot)$ of the $S$-matrix $\tilde S(\cdot)$, the approximate
resonance state defined in Eq. (\ref{multiple_resonance_psi_mu_eqn}) for a
resonance at $z=\mu_j$ is therefore denoted by $\psi_{\mu_j}^\Sigma$. However,
for certain arguments the exact form of $\Sigma$ is irrelevant and it
is useful to define the notion of an $n$'th order approximate resonance state:
\begin{definition}[$\mathbf n$'th order approximate resonance state]
If the number of poles of $\mathcal S(\cdot)$ entering into the definition of
the approximate resonance state $\psi_{\mu_j}^\Sigma$ in Eq.
(\ref{multiple_resonance_psi_mu_eqn}), not including $\mu_j$ itself, is $n$ we
say that $\psi_{\mu_j}^\Sigma$ is an $n$'th order approximate resonance state
for the resonance at $z=\mu_j$. In particular, regardless of the exact nature of the region
$\Sigma$, the zero'th order approximate resonance state is always defined to be
given by Eq. (\ref{approx_resonance_state_explicit_eqn}) with $\mu=\mu_j$ and
is denoted by $\psi_{\mu_j}^{(0)}$.
\end{definition}
\par{\bf Remark:} Note that in general there are many choices of the $n$ resonance
poles (different than $\mu_j$) included in the construction of what we
call an $n$'th order approximation $\psi_{\mu_j}^{(n)}$.
In cases that the nature of the region $\Sigma$ is
irrelevant and only the order of the approximate resonance state is significant
we replace the notation $\psi_{\mu_j}^\Sigma$ by $\psi_{\mu_j}^{(n)}$, where
$n$ is the order of the approximate resonance state considered.
\smallskip
\par The semigroup decomposition and approximate resonance states for the
multi--resonance case possess some interesting properties. For example, we have
\begin{equation*}
 (\psi_{\mu_j}^\Sigma,\psi_{\mu_k}^\Sigma)_{\mathcal H_{ac}(\mathbf H)}
 =(\psi_{\mu_j}^{(0)},\psi_{\mu_k}^{(0)})_{\mathcal H_{ac}(\mathbf H)}
 =\int_{\mathbb R^+} dE \frac{1}{E-\mu_j} \frac{1}{E-\overline\mu_k}
\end{equation*}
and, in particular
\begin{equation}
\label{psi_mu_norm_eqn}
 \Vert\psi_{\mu_j}^\Sigma\Vert^2_{\mathcal H_{ac}(\mathbf H)}
 =\Vert\psi_{\mu_j}^{(n)}\Vert^2_{\mathcal H_{ac}(\mathbf H)}
 =\Vert\psi_{\mu_j}^{(0)}\Vert^2_{\mathcal H_{ac}(\mathbf H)}
 =\int_{\mathbb R^+} dE\, \frac{1}{\vert E-\mu_j\vert^2}\,.
\end{equation}
We see that, although the definition of $\psi_{\mu_i}^\Sigma$ in 
Eq. (\ref{multiple_resonance_psi_mu_eqn}) depends on all of
the poles $\{\mu_j\}_{j=1,\ldots,n}\subset\Sigma\cap\mathbb C^-$, the scalar
product of $\psi_{\mu_i}^\Sigma$ and $\psi_{\mu_j}^\Sigma$ depends only on
$\mu_i$ and $\mu_j$. In fact, if $\mathcal S(\cdot)$ can be extended to a
meromorphic function in a region $\Sigma'\supset\Sigma$ (we keep the notation
$\mathcal S(\cdot)$ for the extended function) and $\mathcal S(\cdot)$ has
now $m>n$ simple poles in $\Sigma'\cap\mathbb C^-$ we may calculate approximate
resonance states of order $m-1$ for all resonances in $\Sigma'$ according to
Eq. (\ref{multiple_resonance_psi_mu_eqn}). However, for $\mu_j,\mu_k\in\Sigma$
we would still have $(\psi_{\mu_j}^{\Sigma'},
\psi_{\mu_k}^{\Sigma'})_{\mathcal H_{ac}(\mathbf H)}=
(\psi_{\mu_j}^{\Sigma},
\psi_{\mu_k}^{\Sigma})_{\mathcal H_{ac}(\mathbf H)}$ i.e., scalar products
(and norms) are independent of the order of the approximate states when we
enlarge the region $\Sigma$. 
In particular we have $\Vert\psi_{\mu_j}^\Sigma\Vert_{\mathcal H_{ac}(\mathbf
H)}=\Vert\psi_{\mu_j}^{(0)}\Vert_{\mathcal H_{ac}(\mathbf H)}$ for every region
$\Sigma$ containing $\mu_j$.
\par An interesting question is whether the peculiar properties of scalar
products and norms of the approximate resonance states mentioned above
characterize also the time evolution of these states. We shall see below
that, at least partially, the answer to this question is positive. For this
we consider one of the basic notions associated with resonance evolution, i.e.,
that of the survival amplitude
\begin{equation}
\label{survival_amplitude_eqn}
  A_{\psi^\Sigma_{\mu_j}}(t)\stackrel{\mathrm{def}}{=}
  \frac{(\psi^\Sigma_{\mu_j}, \mathbf U(t)\psi^\Sigma_{\mu_j})_{\mathcal H_{ac}(\mathbf H)}}
  {(\psi^\Sigma_{\mu_j},\psi^\Sigma_{\mu_j})_{\mathcal H_{ac}(\mathbf H)}}\, .
\end{equation}
Making use of Eq. (\ref{multiple_resonance_psi_mu_eqn}) and (\ref{psi_mu_norm_eqn})
we get a simple expression for this quantity
\begin{equation*}
 A_{\psi^{(0)}_{\mu_j}}(t)=\Vert\psi^{(0)}_{\mu_j}\Vert^{-2}_{\mathcal H_{ac}(\mathbf H)}
 \int_{\mathbb R^+}\frac{1}{\vert E-\mu_j\vert^2}e^{-iE t},\quad t\geq 0
\end{equation*}
where $\Vert\psi^{(0)}_{\mu_j}\Vert$ is given in Eq. (\ref{psi_mu_norm_eqn}). Again,
we see that the expression for the survival amplitude for the approximate
resonance state $\psi^\Sigma_{\mu_j}$ depends only on the pole at $z=\mu_j$ and has
the same form as for a single resonance. This suggests that the semigroup
decompostion of the survival amplitude for the multiple resonance case is
similar to that of a single resonance. When combined with an important
characterization of approximate resonance states in the form of Lemma 
\ref{psi_mu_to_hardy_resonance_lemma} below, such considerations lead
to the following useful \emph{a priori} estimate on the size of the
background term in the semigroup decomposition of the survival amplitude:
\begin{theorem}
\label{background_estimate_thm}
Let $A_{\psi^\Sigma_{\mu_j}}(t)$, $j=1,\ldots,n$ be the survival 
amplitude defined in Eq. (\ref{survival_amplitude_eqn}) and let the background term
$R_{\mu_j}(t)$ be defined by the relation 
\begin{equation}
\label{survival_amplitude_decomp_eqn}
 A_{\psi^\Sigma_{\mu_j}}(t)=R_{\mu_j}(t)+e^{-i\mu_j t},\quad t\geq 0
 \, .
\end{equation}
Then we have
\begin{equation}
\label{background_estimate_eqn}
 \vert R_{\mu_j}(t)\vert
 \leq \left(\frac{\Vert x_{\mu_j}\Vert^4_{H^2_+(\mathbb R)}}
 {\Vert\psi_{\mu_j}^{(0)}\Vert^4_{\mathcal H_{ac}(\mathbf H)}}-1\right)^{1/2},
 \quad t\geq 0\,.
\end{equation}
where $x_{\mu_j}(\sigma)=(\sigma-\mu_j)^{-1}$ is the Hardy space
resonance state and $\psi^{(0)}_{\mu_j}\in\mathcal H_{ac}(\mathbf H)$ is
the zero'th order approximate resonance state corresponding to the resonance at $z=\mu_j$.
\hfill$\square$
\end{theorem}
\par{\bf Proof:} We first have
\begin{proposition}
\label{psi_mu_survival_prop}
For $j=1,\dots,n$, let $\psi_{\mu_j}^\Sigma$ be defined by
Eq. (\ref{multiple_resonance_psi_mu_eqn}) and let $A_{\psi_{\mu_j}^\Sigma}(t)$
be the survival amplitude defined in Eq. (\ref{survival_amplitude_eqn}). Then
\begin{equation}
\label{psi_mu_survival_decomp_eqn}
 A_{\psi_{\mu_j}^\Sigma}(t)
 =\Vert\psi_{\mu_j}^{(0)}\Vert^{-2}_{\mathcal H_{ac}(\mathbf H)}
 (x_{\mu_j},u(t)\mathcal B_{\mu_j}\theta^*
 \tilde S'_{\mu_j}(\theta^*)^{-1}\psi_{\mu_j,in}^{(0)})_{H^2_+(\mathbb R)}
 +e^{-i\mu_j t},\quad t\geq 0\,.
\end{equation}
where $x_{\mu_j}\in H^2_+(\mathbb R)$ is the Hardy space resonance
state and $\psi^{(0)}_{\mu_j}\in\mathcal H_{ac}(\mathbf H)$ is
the zero'th order approximate resonance state corresponding to the
pole at $\mu_j$,
and $\tilde S'_{\mu_j}(\cdot)$ is defined as in
Eq. (\ref{s_tilde_factorization_eqn}), i.e.,
\begin{equation*}
 \tilde S'_{\mu_j}(E)=\frac{E-\mu_j}{E-\overline\mu_j}\,\tilde
 S(E)=\tilde B_{\overline\mu_j}(E)\,\tilde S(E)\, .
\end{equation*}
\hfill$\square$
\end{proposition}
\par{\bf Proof of Proposition \ref{psi_mu_survival_prop}:} We need first the
following easily proved, but important, lemma
\begin{lemma}
\label{psi_mu_to_hardy_resonance_lemma}
For $j=1,\dots,n$, let $\psi^\Sigma_{\mu_j}$ be defined by
Eq. (\ref{multiple_resonance_psi_mu_eqn}). Define
\begin{equation*}
 \mathcal B_{\mu_1\ldots\hat\mu_k\ldots\mu_n}(z)\stackrel{\mathrm{def}}{=}
 \prod_{\substack{i=1\\ i\not= k}}^{n}\frac {z-\overline\mu_i}{z-\mu_i}
\end{equation*}
where $\{\mu_j\}_{j=1,\ldots,n}$ are the poles of $\mathcal S(\cdot)$
in $\Sigma$, and let $x_{\mu_j}(\sigma)=(\sigma-\mu_j)^{-1}$. Then we have
\begin{equation*}
 \psi^\Sigma_{\mu_j}=\hat\Omega_+^* \mathcal B_{\mu_1\ldots\hat\mu_j\ldots\mu_n}\,
 x_{\mu_j}\,.
\end{equation*}
In particular $\psi_{\mu_j}^{(0)}=\hat\Omega_+^* x_{\mu_j}$. \hfill$\square$
\end{lemma}
\par{\bf Proof of Lemma \ref{psi_mu_to_hardy_resonance_lemma}:} 
It is proved in Ref. 19 that, if $\mathbf\Omega^\pm$ are the M\o ller
wave operators, $U: \mathcal H_{ac}(\mathbf H_0)\mapsto L^2(\mathbb
R^+)$ the mapping to the energy representation for
$\mathbf H_0$ (see Eq. (\ref{l2_s_matrix}) above) and $\theta^*$ the
map given in Lemma \ref{theta_star_lemma}, then the quasi-affine
nesting maps $\hat\Omega_\pm$ are given by $\hat\Omega_\pm=\theta^*
U(\mathbf\Omega^\mp)^*$, hence we have
$\hat\Omega_+^*=\mathbf\Omega^- U^*\theta$. Furthermore, by the definition of
$\theta$ we have
\begin{equation}
\label{hardy_resonance_to_l2_eqn}
 (\theta\mathcal B_{\mu_1\ldots\hat\mu_j\ldots\mu_n}x_{\mu_j})(E)
 =\prod_{\substack{i=1\\ i\not=j}}^n\frac{E-\overline\mu_i}{E-\mu_i}\,
 \frac{1}{E-\mu_j},\quad E\in\mathbb R^+\,.
\end{equation}
Moreover, according to Eq. (\ref{outgoing_lipmann_schwinger_eqn}) for every
$g\in L^2(\mathbb R^+)$ we have
\begin{equation}
\label{l2_to_hac_eqn}
 \mathbf\Omega^- U^* g=\int_{\mathbb R^+}dE\,\vert E^-\rangle\, g(E)\,.
\end{equation}
Applying Eq. (\ref{l2_to_hac_eqn}) with
$g=\theta\mathcal B_{\mu_1\ldots\hat\mu_j\ldots\mu_n}x_{\mu_j}$ and comparing
with Eq. (\ref{multiple_resonance_psi_mu_eqn}) proves the lemma 
\par{}\hfill$\blacksquare$
\par\bigskip
Note that by Lemma \ref{psi_mu_to_hardy_resonance_lemma} we have
$(\hat\Omega_+^*)^{-1}\psi^\Sigma_{\mu_j}=\mathcal B_{\mu_1\ldots\hat\mu_j\ldots\mu_n}
x_{\mu_j}$. Hence by Eq. (\ref{sem_dec_multi_resonance_eqn}) we get
\begin{multline}
\label{multi_resonance_survival_eqn}
 A_{\psi^\Sigma_{\mu_j}}(t)=\Vert\psi^\Sigma_{\mu_j}\Vert^{-2}_{\mathcal H_{ac}(\mathbf H)}
 (\psi^\Sigma_{\mu_j},\mathbf U(t)\psi^\Sigma_{\mu_j})_{\mathcal H_{ac}(\mathbf H)}=\\
 =\Vert\psi^\Sigma_{\mu_j}\Vert^{-2}_{\mathcal H_{ac}(\mathbf H)}
 (\mathcal B_{\mu_1\ldots\hat\mu_j\ldots\mu_n}x_{\mu_j}
 ,u(t)\mathcal B_{\mu_1\ldots\mu_n}\theta^*
 \tilde S'(\theta^*)^{-1}\psi^\Sigma_{\mu_j,in})_{H^2_+(\mathbb R)}\\
 +\sum_{k=1}^n\frac{\vert{\rm Im}\,\mu_k\vert}{\pi}\,
 \prod_{\substack{i=1\\ i\not=k}}^n\frac{\mu_k-\overline\mu_i}{\mu_k-\mu_i}
 \frac{(\psi^\Sigma_{\mu_k},\psi^\Sigma_{\mu_j})_{\mathcal H_{ac}(\mathbf H)}}
 {\Vert\psi^\Sigma_{\mu_j}\Vert^2_{\mathcal H_{ac}(\mathbf H)}}
 (\mathcal B_{\mu_1\ldots\hat\mu_j\ldots\mu_n}x_{\mu_j}
 ,x_{\mu_k})_{H^2_+(\mathbb R)}e^{-i\mu_k t}\\
\end{multline}
In the second term on the r.h.s. of
Eq. (\ref{multi_resonance_survival_eqn}) we first 
separate the term with $k=j$ and get 
\begin{multline*}
 \sum_{k=1}^n\frac{\vert{\rm Im}\,\mu_k\vert}{\pi}\,
 \prod_{\substack{i=1\\ i\not=k}}^n\frac{\mu_k-\overline\mu_i}{\mu_k-\mu_i}
 \frac{(\psi^\Sigma_{\mu_k},\psi^\Sigma_{\mu_j})_{\mathcal H_{ac}(\mathbf H)}}
 {\Vert\psi^\Sigma_{\mu_j}\Vert^2_{\mathcal H_{ac}(\mathbf H)}}
 (\mathcal B_{\mu_1\ldots\hat\mu_j\ldots\mu_n}x_{\mu_j}
 ,x_{\mu_k})_{H^2_+(\mathbb R)}e^{-i\mu_k t}=\\
 =\sum_{\substack{k=1\\k\not=j}}^n\frac{\vert{\rm Im}\,\mu_k\vert}{\pi}\,
 \prod_{\substack{i=1\\ i\not=k}}^n\frac{\mu_k-\overline\mu_i}{\mu_k-\mu_i}
 \frac{(\psi^\Sigma_{\mu_k},\psi^\Sigma_{\mu_j})_{\mathcal H_{ac}(\mathbf H)}}
 {\Vert\psi^\Sigma_{\mu_j}\Vert^2_{\mathcal H_{ac}(\mathbf H)}}
 (\mathcal B_{\mu_1\ldots\hat\mu_j\ldots\mu_n}x_{\mu_j}
 ,x_{\mu_k})_{H^2_+(\mathbb R)}e^{-i\mu_k t}+e^{-i\mu_j t}\\
 ,\quad t\geq 0
\end{multline*}
Here, use has been made of Eq. (\ref{x_mu_norm_eqn}) below. 
The above expression can be further simplified since  
$x_{\mu_k}\in\hat{\mathcal K}_{\mu_k}
=H^2_+(\mathbb R)\ominus\mathcal B_{\mu_k}(\cdot)H^2_+(\mathbb R)$ implies
that for $k\not=j$ we have 
\begin{equation*}
 (\mathcal B_{\mu_1\ldots\hat\mu_j\ldots\mu_n}x_{\mu_j},
 x_{\mu_k})_{H^2_+(\mathbb R)}=(\mathcal B_{\mu_k}
 \mathcal B_{\mu_1\ldots\hat\mu_j\ldots\hat\mu_k\ldots\mu_n}x_{\mu_j},
 x_{\mu_k})_{H^2_+(\mathbb R)}=0
\end{equation*}
where
\begin{equation*}
 \mathcal B_{\mu_1\ldots\hat\mu_j\ldots\hat\mu_k\ldots\mu_n}(z)\stackrel{\mathrm{def}}{=}
 \prod_{\substack{i=1\\ i\not=k,j}}^{n}\frac
 {z-\overline\mu_i}{z-\mu_i}\, .
\end{equation*}
The first term on the r.h.s. of Eq. (\ref{multi_resonance_survival_eqn}) 
can also be simplified. We have
\begin{multline*}
 (\mathcal B_{\mu_1\ldots\hat\mu_j\ldots\mu_n}x_{\mu_j}
 ,u(t)\mathcal B_{\mu_1\ldots\mu_n}\theta^*
 \tilde S'(\theta^*)^{-1}\psi^\Sigma_{\mu_j,in})_{H^2_+(\mathbb R)}=\\
 =(u(-t)\mathcal B_{\mu_1\ldots\hat\mu_j\ldots\mu_n}x_{\mu_j}
 ,\mathcal B_{\mu_1\ldots\mu_n}\theta^*
 \tilde S'(\theta^*)^{-1}\psi^\Sigma_{\mu_j,in})_{H^2_+(\mathbb R)}=\\
 =(\mathcal B_{\mu_1\ldots\hat\mu_j\ldots\mu_n}u(-t)x_{\mu_j}
 ,\mathcal B_{\mu_1\ldots\mu_n}\theta^*
 \tilde S'(\theta^*)^{-1}\psi^\Sigma_{\mu_j,in})_{H^2_+(\mathbb R)}=\\
 =(x_{\mu_j},u(t)\mathcal B_{\mu_j}\theta^*
 \tilde S'(\theta^*)^{-1}\psi^\Sigma_{\mu_j,in})_{H^2_+(\mathbb R)}
 ,\quad t\geq 0
\end{multline*}
Moreover, using Lemma \ref{psi_mu_to_hardy_resonance_lemma} we find that
\begin{multline*}
 \theta^*\tilde S'(\theta^*)^{-1}\psi^\Sigma_{\mu_j,in}
 =\theta^*\tilde S'(\theta^*)^{-1}
 \hat\Omega_-\hat\Omega_+^*\mathcal B_{\mu_1\ldots\hat\mu_j\ldots\mu_n}
 x_{\mu_j}=\\
 =\theta^*\tilde S'(\theta^*)^{-1}\theta^*\tilde S^*\theta
 \mathcal B_{\mu_1\ldots\hat\mu_j\ldots\mu_n}x_{\mu_j}
 =\theta^* \overline{\tilde B_{\mu_1\ldots\mu_n}}\theta
 \mathcal B_{\mu_1\ldots\hat\mu_j\ldots\mu_n}x_{\mu_j}=\\
 =\theta^*\overline{\tilde B_{\mu_j}}
 \theta x_{\mu_j}
 =\theta^*\tilde B_{\overline\mu_j}\tilde S(\theta^*)^{-1}
 \hat\Omega_-\hat\Omega_+^* x_{\mu_j}
 =\theta^* \tilde S'_{\mu_j}(\theta^*)^{-1}\psi^0_{\mu_j,in}\,.\\
\end{multline*}
Recalling that Eq. (\ref{psi_mu_norm_eqn}) implies that 
$\Vert\psi_{\mu_j}^\Sigma\Vert_{\mathcal H_{ac}(\mathbf H)}
=\Vert\psi_{\mu_j}^{(0)}\Vert_{\mathcal H_{ac}(\mathbf H)}$ the proof of
Proposition \ref{psi_mu_survival_prop} is complete. \hfill$\blacksquare$
\bigskip
\par From Proposition \ref{psi_mu_survival_prop} we see that, independent
of the region $\Sigma$, the semigroup decomposition of the
survival amplitude depends only on the zero'th order approximate resonance
state. Comparison of Eq. (\ref{psi_mu_survival_decomp_eqn}) and
Eq. (\ref{survival_amplitude_decomp_eqn}) gives 
\begin{equation}
\label{survival_background_term_eqn}
 R_{\mu_j}(t)=
 \Vert\psi_{\mu_j}^{(0)}\Vert^{-2}_{\mathcal H_{ac}(\mathbf H)}
 (x_{\mu_j},u(t)\mathcal B_{\mu_j}\theta^*
 \tilde S'_{\mu_j}(\theta^*)^{-1}\psi_{\mu_j,in}^{(0)})_{H^2_+(\mathbb R)}
 ,\quad t\geq 0\, ,
\end{equation}
This expression for $R_{\mu_j}(t)$ is identical to the zero'th order
background term we would get from
Eq. (\ref{semigroup_decomp_approx_state_eqn}) with
$f=g=\psi^{(0)}_{\mu_j}$. We now exploit this fact to obtain the
desired estimate in Theorem 
\ref{background_estimate_thm}. Applying Theorem 
\ref{approx_resonance_state_thm} to the
zero'th order approximate resonance state $\psi^{(0)}_{\mu_j}$ we obtain
\begin{equation*}
 \psi^{(0)}_{\mu_j,out}=\hat\Omega_+\psi^{(0)}_{\mu_j}=
 \mathcal B_{\mu_j}\theta^*\tilde S'_{\mu_j}(\theta^*)^{-1}\psi^{(0)}_{\mu_j,in}
 +\frac{\vert{\rm Im}\,\mu_j\vert}{\pi}
 \Vert\psi^{(0)}_{\mu_j}\Vert^2_{\mathcal H_{ac}(\mathbf H)}x_{\mu_j}\,.
\end{equation*}
Now, since both $\hat\Omega_+$ and $\hat\Omega_+^*$ are contractive we note
that Lemma \ref{psi_mu_to_hardy_resonance_lemma} implies that
$\Vert\psi^{(0)}_{\mu_j,out}\Vert_{H^2_+(\mathbb R)}
=\Vert\hat\Omega_+\hat\Omega_+^* x_{\mu_j}\Vert_{H^2_+(\mathbb R)}\leq
\Vert x_{\mu_j}\Vert_{H^2_+(\mathbb R)}$. In addition in $H^2_+(\mathbb R)$ we
have $x_{\mu_j}\perp \mathcal B_{\mu_j}H^2_+(\mathbb R)$. Therefore, 
\begin{equation*}
 \Vert\mathcal B_{\mu_j}\theta^*\tilde S'_{\mu_j}(\theta^*)^{-1}\psi^{(0)}_{\mu_j,in}
 \Vert^2_{H^2_+(\mathbb R)}+\frac{\vert{\rm Im}\,\mu_j\vert^2}{\pi^2}
 \Vert\psi^{(0)}_{\mu_j}\Vert^4_{\mathcal H_{ac}(\mathbf H)}
 \Vert x_{\mu_j}\Vert^2_{H^2_+(\mathbb R)}
 \leq \Vert x_{\mu_j}\Vert^2_{H^2_+(\mathbb R)}\, .
\end{equation*}
It is easy to verify that 
\begin{equation}
\label{x_mu_norm_eqn}
 \Vert x_{\mu_j}\Vert^2_{H^2_+(\mathbb R)}=\int_{-\infty}^{\infty}\, 
 \frac{1}{\vert\sigma-\mu_j\vert^2}\,d\sigma=\frac{\pi}{\vert{\rm Im}\,\mu_j\vert},
\end{equation}
hence the inequality above can be written in the form 
\begin{equation}
\label{basic_background_estimate_eqn}
 \Vert\mathcal B_{\mu_j}\theta^*\tilde S'_{\mu_j}(\theta^*)^{-1}\psi^{(0)}_{\mu_j,in}
 \Vert^2_{H^2_+(\mathbb R)}\leq\left(\Vert x_\mu\Vert^4_{H^2_+(\mathbb R)}
 -\Vert\psi^{(0)}_{\mu_j}\Vert^4_{\mathcal H_{ac}(\mathbf H)}\right)
 \Vert x_{\mu_j}\Vert^{-2}_{H^2_+(\mathbb R)}\, .
\end{equation}
Applying the Schwartz inequality to the r.h.s. of Eq.
(\ref{survival_background_term_eqn}) and using the bound from Eq.
(\ref{basic_background_estimate_eqn}) we get the estimate in Eq.
(\ref{background_estimate_eqn}).\hfill$\blacksquare$\break
\bigskip
\par As mentioned above the background term cannot be identically
zero. Hence deviations from exponential decay of the survival probability
are to be expected. In fact, it is easy to verify that the survival probability
behaves for short times as $\vert A_{\psi_{\mu_j}}(t)\vert^2=1-O(t^2)$.
Note that Eq. (\ref{survival_amplitude_decomp_eqn}) implies that at $t=0$ we
must have $R_{\mu_j}(0)=0$. This is also seen from
Eq. (\ref{survival_background_term_eqn}), since for $t=0$ we have
$u(0)=1$ and $x_{\mu_j}\perp\mathcal B_{\mu_j}(\cdot)H^2_+(\mathbb
R)$. Deviations from exponential decay are then due to the
fact that $x_{\mu_j}\not\perp u(t)\mathcal B_{\mu_j}H^2_+(\mathbb
R)$ for $t>0$.

\section{Example: Scattering from square barrier potential}
\label{example}
\par In this section we apply the results of the previous two sections
to a simple one dimensional model with a square barrier potential. 
Although simple, this model provides a good
illustration for the various results obtained above. In particular, we
present numerical calculations of approximate resonance states of
various orders accompanied with plots of the time evolution of the
corresponding survival amplitudes and estimates of the size of the background
term following from Theorem \ref{background_estimate_thm}.
\par The model we consider is a Schr\"odinger equation in one
spatial dimension on the half-line $\mathbb R^+$ with a square barrier
potential. Thus we consider the free Hamiltonian 
$\mathbf H_0=-\partial^2_x$ acting on $L^2(\mathbb R^+)$ (where
$\mathbf H_0$ is defined as a self-adjoint extension to $L^2(\mathbb
R^+)$ from the original domain of definition 
$D(-\partial^2_x)=\{\phi(x)\in W_2^2(\mathbb R^+)\mid \phi(x)=0\}$) 
and the full Hamiltonian is given by
$\mathbf H=\mathbf H_0+\mathbf V$ where $\mathbf V$ is a 
multiplicative operator $(\mathbf Vf)(x)=V(x)f(x)$ with  
\begin{equation*}
 V(x)= \begin{cases} 0, & 0<x<a\\ V_0, & a\leq x\leq b\\ 0, & b<x,\\
 \end{cases}
\end{equation*}
where $b>a>0$ and we take $V_0>0$. In this case there are no bound
state solutions of the eigenvalue problem for $\mathbf H$ and we have
$\sigma(\mathbf H)=\sigma_{ac}(\mathbf H)=\mathbb R^+$. In order to
find the scattering states, calculate the $S$-matrix and finally the 
approximate resonance states for this problem one solves the eigenvalue problem 
\begin{equation*}
 -\partial_x^2\psi_E(x)+V(x)\psi_E(x)=E\psi_E(x),\quad E\in\mathbb R^+
\end{equation*}
for the continuous spectrum generalized eigenfunctions
$\psi_E(x)$. Imposing boundary conditions one finds
that
\begin{equation}
\label{cont_spec_eigenstates_eqn}
 \psi_E(x)=\begin{cases} \alpha_1(k)\sin kx, & 0<x\leq a\\
 \alpha_2(k)e^{ik'x}+\beta_2(k)e^{-ik' x}, & a<x<b\\
 \alpha_3(k)e^{ikx}+\beta_3(k)e^{-ikx}, & b\leq x\\
 \end{cases}
\end{equation}
where $k=E^{1/2}$ and $k'=\sqrt{E-V_0}$ for $E\geq V_0>0$ or
$k'=i\sqrt{V_0-E}$ for $V_0>E>0$. The coefficients in Eq. (\ref{cont_spec_eigenstates_eqn}) 
are given by${}^{32}$
\begin{eqnarray}
\label{coefficients_cont_spec_states_eqn}
 \alpha_2(k) &=& \frac{1}{2}e^{-ik'a}\left[\sin ka+\frac{k}{ik'}\cos ka\right]\alpha_1(k)\nonumber\\
 \beta_2(k) &=& \frac{1}{2}e^{ik'a}\left[\sin ka-\frac{k}{ik'}\cos ka\right]\alpha_1(k)\nonumber\\
 \alpha_3(k) &= &\frac{1}{4}e^{-ikb}\left[(1+k'/k)e^{ik'(b-a)}(\sin
     ka+\frac{k}{ik'}\cos ka)\right.\\
 & & +\left. (1-k'/k)e^{-ik'(b-a)}(\sin
     ka-\frac{k}{ik'}\cos ka)\right]\alpha_1(k)\nonumber\\
 \beta_3(k) &=& \frac{1}{4}e^{ikb}\left[(1-k'/k)e^{ik'(b-a)}(\sin
     ka+\frac{k}{ik'}\cos ka)\right.\nonumber \\
 & & +\left. (1+k'/k)e^{-ik'(b-a)}(\sin
     ka-\frac{k}{ik'}\cos ka)\right]\alpha_1(k)\nonumber
\end{eqnarray}
with $\alpha_1(k)$ to be determined by normalization conditions (see below). 
\par Given the full set of solutions $\{\psi_E(x)\}_{E\in\mathbb R^+}$
for the continuous spectrum it is easy to find the sets
$\{\psi^\pm_E(x)\}_{E\in\mathbb R^+}$ of solutions of the
Lippmann-Schwinger equation corresponding to incoming and outgoing
asymptotic conditions. Using Dirac's notation we have${}^{32}$
\begin{eqnarray}
\label{lipmann_schwinger_states_eqn}
 \langle x\vert E^+\rangle &\equiv &
 \psi^+_E(x)=\frac{-1}{2i}\frac{\psi_E(x)}{\beta_3(k)}\\
 \langle x\vert E^-\rangle &\equiv
 &\psi^-_E(x)=\frac{1}{2i}\frac{\psi_E(x)}{\alpha_3(k)}\, .\nonumber
\end{eqnarray}
\begin{figure}[t]
\setlength{\unitlength}{1cm}
\setlength{\fboxsep}{0cm}
\hskip 8cm
 \begin{minipage}[t]{5.0cm}
   \mbox{\scalebox{1.0}{\includegraphics[height=5.5cm,width=7.0cm]{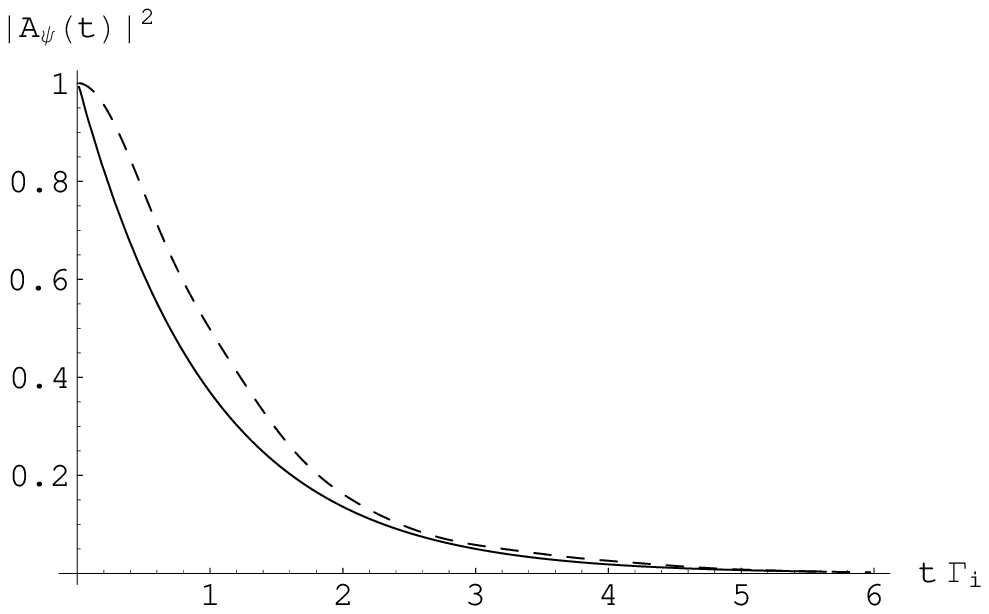}}}
   \parbox{7cm}{\caption[fig3]{\label{survival_prob_fig} \scriptsize
       survival amplitudes: Solid line - $\vert
      A_{\psi_{\mu_1}}(t)\vert^2$, dashed line - $\vert
      A_{\psi_{\mu^{\prime}_3}}(t)\vert^2$.Insert shows short time
    behaviour of $\vert A_{\psi_{\mu_1}}(t)\vert^2$.}}
 \end{minipage}
\hskip -3.2cm
\begin{minipage}[t]{5.0cm}
   \raisebox{1.5cm}{\mbox{\scalebox{1.0}{\includegraphics[height=4.2cm,width=5.0cm]{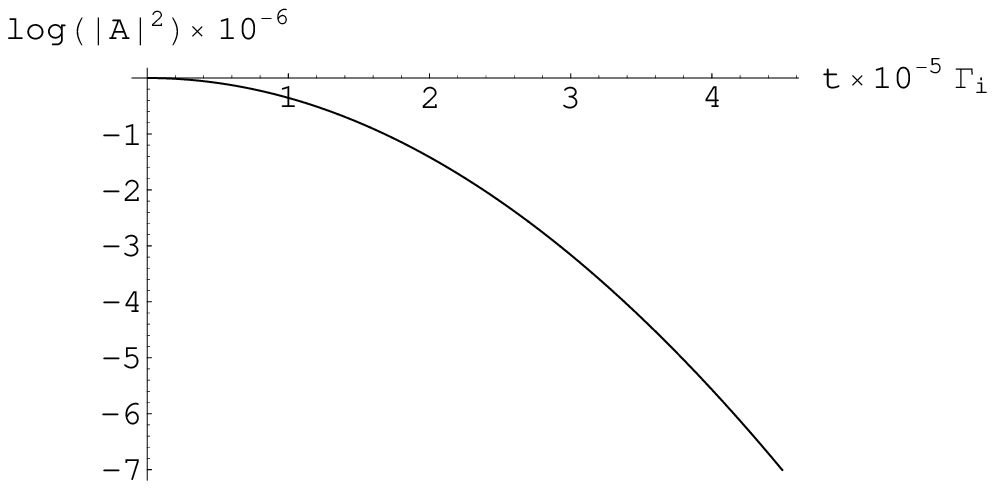}}}}
 \end{minipage}
\setlength{\unitlength}{1cm}
\vskip -7.5cm
\hskip 1cm
\begin{minipage}[t]{4.5cm}
   \mbox{\scalebox{1.0}{\includegraphics[height=12.6cm ,width=6.5cm]{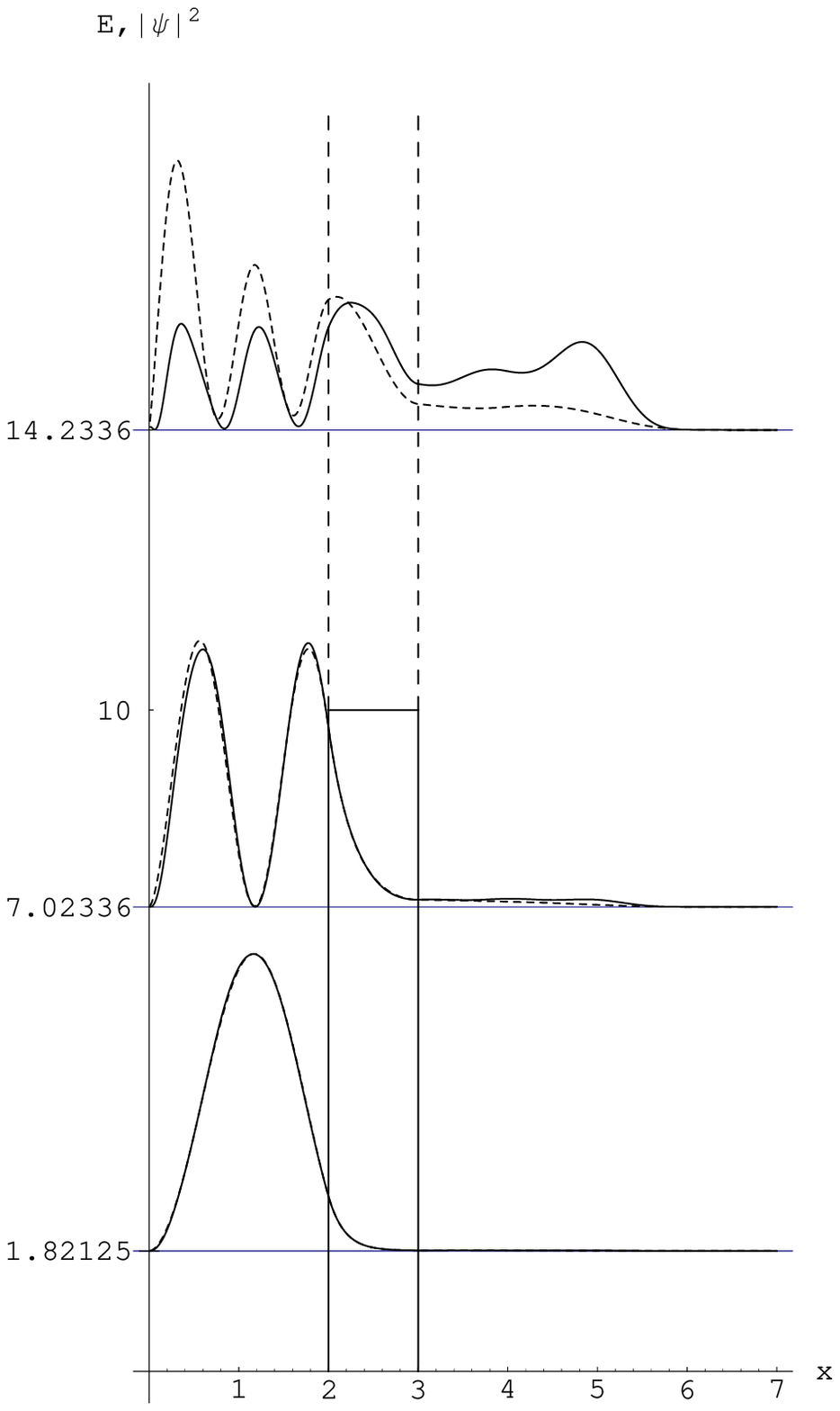}}\par}
   \parbox{6.5cm}{\caption[fig2]{\label{prob_density_fig} \scriptsize Probability
     densities of approximate resonance states 
     $\psi^{(n)}_{\mu_j}$, $j=1,2,3$ for $a=2$,
     $b=3$, $V_0=10$. Dashed lines - $\vert\psi^{(0)}_{\mu_j}(x)\vert^2$,
     $j=1,2,3$. Solid lines - $\vert\psi^{(9)}_{\mu_j}(x)\vert^2$, $j=1,2,3$.}}
\end{minipage}
\hskip 2.5cm
 \begin{minipage}[t]{5.0cm}
    \mbox{\scalebox{1.0}{\includegraphics[height=5.0cm,width=7.0cm]{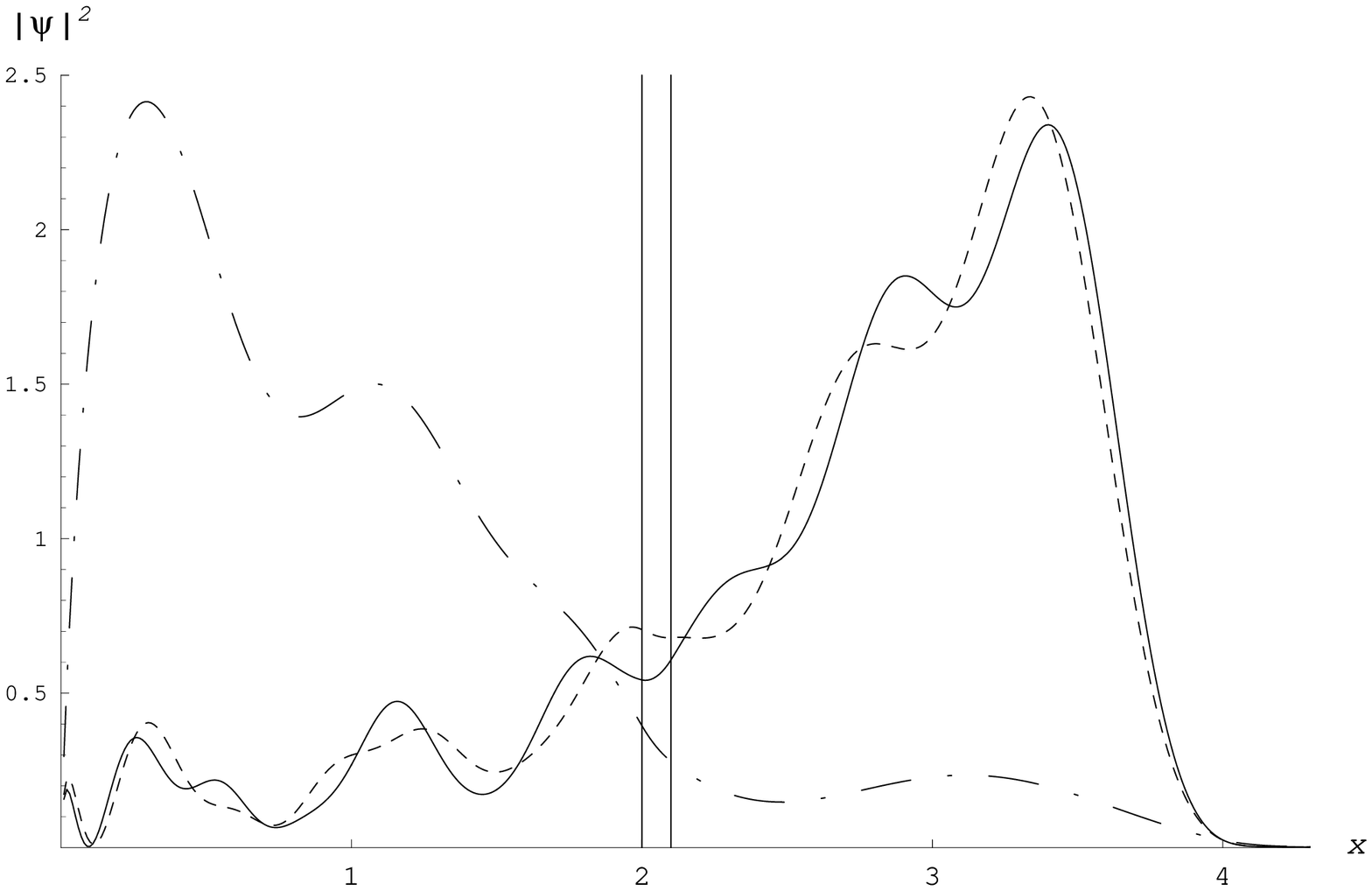}}\par}
    \parbox{7cm}{\caption[fig4]{\label{high_order_state_fig} \scriptsize Probability
     densities of the approximate resonance states for $\mu'_3=17.4652-i4.4029$. Dot
     dashed line - $\vert\psi^{(0)}_{\mu'_3}(x)\vert^2$,
     dashed line - $\vert\psi^{(8)}_{\mu'_3}(x)\vert^2$, solid line - 
     $\vert\psi^{(9)}_{\mu'_3}(x)\vert^2$.}}
 \end{minipage}
\end{figure}
The normalization conditions for the Lippmann-Schwinger states in
Eq. (\ref{lipmann_schwinger_states_eqn}) determines $\alpha_1(k)$ with the
result $\alpha_1(k)=(2\pi k)^{-1/2}$.  
In the energy representation the $S$-matrix is then given by${}^{32}$
\begin{equation*}
 \tilde S(E)=-\frac{\alpha_3(k)}{\beta_3(k)},\quad k=E^{1/2}
\end{equation*}
We now have the ingredients for the calculation of the scattering
resonances and the corresponding approximate resonance states. Note 
first that $\alpha_3(k)$ and $\beta_3(k)$ can be extended to
analytic functions in the complex $k$ plane and as a result the
poles of the analytic continuation of $\tilde S(E)$ to the lower 
half-plane (i.e., across the square root cut along the positive real
axis) are identified with zeros of the function $\beta_3(k)$. For
a resonance at a point $z=\mu_j$ in the lower half-plane below the
positive real axis we set $\mu_j=E_{\mu_j}-i\Gamma_{\mu_j}/2$, with
$E_{\mu_j}$ being the resonance energy and $\Gamma_{\mu_j}$ its width.
\par Using Eq. (\ref{approx_resonance_state_explicit_eqn}) and the
expression for the outgoing Lippmann-Schwinger eigenfunctions
$\langle x\vert E^-\rangle$, Eq. (\ref{cont_spec_eigenstates_eqn}), 
(\ref{coefficients_cont_spec_states_eqn}) and
(\ref{lipmann_schwinger_states_eqn}), the zero'th order approximate
resonance states for the square barrier problem can be calculated numerically.  
Considering a larger number of resonance poles we are able to
calculate higher order approximate resonance states using
Eq. (\ref{multiple_resonance_psi_mu_eqn}). As an example we consider
the three lowest energy resonance
poles for barrier parameters $a=2$, $b=3$, $V_0=10$. These poles are located at
$\mu_1=1.8213-i 0.0023$, $\mu_2=7.0237-i 0.0564$ and
$\mu_3=14.2336-i 0.8923$. The zero'th order probability densities
$\vert\psi^{(0)}_{\mu_j}(x)\vert^2$, $j=1,2,3$ for these poles 
are shown as dashed lines in Fig. \ref{prob_density_fig} while the solid
lines on the same figure correspond to the 9'th order probability densities 
$\vert\psi^{(9)}_{\mu_j}(x)\vert^2$, $j=1,2,3$, where the ten lowest
energy resonances $\mu_j$, $j=1,\ldots,10$ are taken into account in
Eq. (\ref{multiple_resonance_psi_mu_eqn}). We observe the significant change in the
probability density profile between the zero'th and 9'th order
approximate resonance states for the resonance $\mu_3$, whose energy
is higher then the barrier's energy, while the lower two states are
essentially unchanged.
Numerical calculations show that approximate resonance states 
converge in $L^2$ norm to a limiting state as a function of the order of
approximation. An example is provided in Fig. \ref{high_order_state_fig} 
which shows the probability density $\vert\psi_{\mu'_3}^{(n)}(x)\vert^2$
for the resonance $\mu'_3=17.4652-i 4.4029$, at barrier parameters
$a=2$, $b=2.1$, $V_0=10$, and for the orders $n=0,8,9$. At present
a rigorous criterion for the rate of convergence of approximate
resonance states as a function of order is not yet established.
\par Turning to a consideration of the time evolution of survival
probabilities for resonances of the square barrier model, we first recall
the fact that the time evolution of the survival probabilities of
higher order approximate resonance states corresponding to the same
resonance pole is independent of the order and is, in fact, identical
to that of the zero'th order state. Bearing this in mind we may omit
in our notation any indication of the region $\Sigma$ or the order $n$ and set 
$A_{\psi_{\mu_j}}(t)\equiv A_{\psi^\Sigma_{\mu_j}}(t)=A_{\psi^{(n)}_{\mu_j}}(t)$. The time
dependence of the survival probablility $\vert A_{\psi_{\mu_1}}(t)\vert^2$
for the states coresponding to the lower resonance in Fig. 
\ref{prob_density_fig} is shown as a solid line in 
Fig. \ref{survival_prob_fig}. The time evolution of 
$\vert A_{\psi_{\mu_1}}(t)\vert^2$ follows closely an exact exponential decay
law with a decay constant $\Gamma_{\mu_1}=2\vert\Im(\mu_1)\vert$.
This behaviour is reflected in the bound $\vert R_{\mu_1}(t)\vert\leq 0.028$
on the size of the background term calculated using Theorem
\ref{background_estimate_thm}. The time development of  
$\vert A_{\psi_{\mu_1}}(t)\vert^2$ deviates from the exponential law
at a very short time scale, as is clearly seen in the insert in
Fig. \ref{survival_prob_fig}. The behaviour of the survival
probability for the other resonances in Fig. \ref{prob_density_fig} (not shown in
Fig. \ref{survival_prob_fig}) is similar. The short time deviations from
exponential decay are related to the known Zeno effect.  
\par The nearly exact exponential decay law of the survival
probability $\vert A_{\psi_{\mu_1}}(t)\vert^2$ is to be contrasted with the
time development of $\vert A_{\psi_{\mu'_3}}(t)\vert^2$ for the states in
Fig. \ref{high_order_state_fig}. The survival probability 
$\vert A_{\psi_{\mu'_3}}(t)\vert^2$ is described by the dashed line in
Fig. \ref{survival_prob_fig}. Deviations from an exponential decay law
in this case are evidently larger. This conforms with the results of
Theorem \ref{background_estimate_thm} which produces the larger bound 
$\vert R_{\mu'_3}(t)\vert\leq 0.422$. 
\section{Summary}
\label{summary}
\par The semigroup decomposition formalism makes use of the
fundamental mathematical theory underlying the structure of the
Lax-Phillips scattering theory, i.e., the functional model for
$C_{\cdot 0}$ contractive semigroups, for the description of the time
evolution of resonances. If the $S$-matrix is meromorphic in a region 
$\Sigma$ and is known to have resonance poles there at points
$z=\mu_1,\ldots,\mu_n\in\Sigma$, the semigroup formalism allows for the association of a 
unique Hilbert space state $\psi_{\mu_j}^\Sigma(x)\in\mathcal
H_{ac}(\mathbf H)$, $j=1,\ldots n$ with each resonance. The states
$\psi_{\mu_j}^\Sigma(x)$, called approximate resonance states, define
the decomposition of matrix elements of the evolution and are
associated with its semigroup part. Theorem
\ref{background_estimate_thm} 
provides an upper bound on the size of the remaining background
term. Depending on one's knowledge of the location of the resonance
poles it is possible to calculate approximate resonance states of
different orders. Numerical calculations show that the sequence of
approximate resonance states appear to converge in $L^2$ norm to a limiting
function as a function of the order. However, rigorous criteria for
the rate of convergence are needed. Another possible course of further
investigation involves the study of relations between known frameworks
for the treatment of the problem of resonances, such as the rigged 
Hilbert space method and the use of dilation analyticity and the
formalism discussed in the present paper.
\bigskip
\par{\Large\bf Acknowledgements}
\smallskip
\par The work of Y. Strauss was partially supported by ISF under Grant
No. 1282/05 and Grant No. 188/02, and by the Center for Advanced
Studies in Mathematics at Ben-Gurion University and the Edmond Landau Center for
research in Mathematical Analysis and related areas, sponsored by the
Minerva Foundation (Germany).
\bigskip
\begin{description}
\small
\item{}${}^1$P.D. Lax and R.S. Phillips, \emph{Scattering
        Theory} (Academic Press, New York, 1967).
\item{}${}^2$V.M. Adamjan, Funkts. Anal. Prilozh. {\bf 10}, 1
      (1976).
\item{}${}^3$J. Cooper and W. Strauss, Indiana Univ. Math. J. {\bf
        34}, 33 (1985).
\item{}${}^4$R. Phillips, Indiana Univ. Math. J. {\bf 33}, 832
      (1984).
\item{}${}^5$C. Foias, J. Funct. Anal. {\bf 19}, 273 (1975).
\item{}${}^6$D.Z. Arov, Soviet Math. Dokl. {\bf 15}. 848 (1974).
\item{}${}^7$J. Sj\" ostrand and M. Sworski, J. Funct. Anal. {\bf
        123}, 336 (1994).
\item{}${}^8$S. Kuzhel, Ukrainian Math. J. {\bf 55}, no. 5, 621
      (2003).
\item{}${}^9$S. Kuzhel, Methods Funct, Anal. Topology {\bf 7}, 13
      (2001).
\item{}${}^{10}$S. Kuzhel and U. Moskalyova, J. Math. Kyoto
      Univ. {\bf 45}, no. 2, 265 (2005).
\item{}${}^{11}$C. Flesia and C. Piron, Helv. Phys. Acta {\bf 57},
      697 (1984).
\item{}${}^{12}$L.P. Horwitz and C. Piron, Helv. Phys. Acta {\bf 66},
      693 (1993).
\item{}${}^{13}$E. Eisenberg and L.P. Horwitz, ``Time,
      irreversibility, and unstabel systems in quantum mechanics'', in
      \emph{Advnces in Chemical Physics}, edited by I. Prigogine and
      S. Rice (Wiley, New York, 1997), Vol. XCIX.
\item{}${}^{14}$Y. Strauss, L.P. Horwitz, and E. Eisenberg,
      J. Math. Phys. {\bf 41}, 8050 (2000).
\item{}${}^{15}$Y. Strauss and L.P. Horwitz, Found. Phys. {\bf 30},
      653 (2000).
\item{}${}^{16}$Y. Strauss and L.P. Horwitz, J. Math. Phys., {\bf
        43}, 2394 (2002).
\item{}${}^{17}$T. Ben Ari and L.P. Horwitz, Phys. Lett. A {\bf
        332}, 168 (2004).
\item{}${}^{18}$M. Reed and B. Simon, \emph{methods of modern
        mathematical physics, Vol. 3, Scattering theory} (Academic
        Press, New York, 1979).
\item{}${}^{19}$Y. Strauss, J. Math. Phys. {\bf 46}, 32104 (2004).
\item{}${}^{20}$Y. Strauss, J. Math. Phys. {\bf 46}, 102109 (2005).
\item{}${}^{21}$B. Sz.-Nagy and C. Foias, \emph{Harmonic Analysis of
        Operators on Hilbert Space} (North-Holland, Amsterdam,
        London, 1970).
\item{}${}^{22}$H. Baumg\" artel, Rep. Math. Phys. {\bf 52}, 295 (2003)
      and errata, Rep. Math. Phys. {\bf 53}, 329 (2004).
\item{}${}^{23}$H. Baumg\" artel, Rev. Math. Phys. {\bf 18}, 61
        (2006).
\item{}${}^{24}$G. Gamow, Z. Phys. {\bf 51}, 204 (1928).
\item{}${}^{25}$C.S. Kubrusly, \emph{An introduction to Models and
        Decompositions in Operator Theory} (Birkhauser, Boston, 1997).
\item{}${}^{26}$M. Rosenblum and J. Rovnyak, \emph{Hardy classes
        and Operator Theory} (Oxford University Press, New York,
        1985).
\item{}${}^{27}$N.K. Nikolski\u{i}, \emph{Treatise on the Shift
        Operator} (Springer-Verlag, New York, 1986).
\item{}${}^{28}$Y. Strauss, Int. J. Theo. Phys. {\bf 42}, 2285
      (2003).
\item{}${}^{29}$K. Hoffman, \emph{Banach Spaces of Analytic
        Functions} (Prentice Hall, Englewood Cliffs, NJ, 1962).
\item{}${}^{30}$P.L. Duren, \emph{Theory of $\mathcal H^p$ Spaces}
      (Academic, New York, London, 1970).
\item{}${}^{31}$C. Van Winter, Trans. Am. Mat. Soc. {\bf 162}, 103
      (1971).
\item{}${}^{32}$R. de la Madrid and M. Gadella, Amer. J. Phys. {\bf
        70}, 626 (2002).
\end{description}

\end{document}